\documentclass[preprint,nofootinbib,12pt,floatfix,aps,prd,superscriptaddress]{revtex4-1}
\usepackage{amsmath,amssymb,amsfonts,amstext,amsbsy,amsopn}
\usepackage{graphicx, color, xcolor, slashed, esint}
\usepackage[dvips]{epsfig}
\usepackage{float, units, textcomp, hyperref}
\usepackage[utf8]{inputenc}
\usepackage{subcaption}
\usepackage[normalem]{ulem}
\newcommand{\beq}{\begin{equation}}
\newcommand{\eeq}{\end{equation}}
\newcommand{\bea}{\begin{eqnarray}}
\newcommand{\eea}{\end{eqnarray}}

\newcommand{\setemeio}{{}^7\!/{}_{\!2}}

\begin{document}
%\title{Cylindrical black holes surrounded by Dekel-Zhao dark matter profile}
%\title{Analytical Solutions for Black Strings and BTZ Black Hole in $(3+1)$ Dimensions}
\title{Cylindrically Symmetric
Black Holes Sourced by Dekel–Zhao Dark Matter}
\author{G. Alencar}
\email{geova@fisica.ufc.br}
\affiliation{Departamento de Física, Universidade Federal do Ceará, Caixa Postal 6030, Campus do Pici, 60455-760 Fortaleza, Ceará, Brazil}
\author{V. H. U. Borralho}
\email{victorborralho@fisica.ufc.br} 
\affiliation{Departamento de Física, Universidade Federal do Ceará, Caixa Postal 6030, Campus do Pici, 60455-760 Fortaleza, Ceará, Brazil}
\author{M. S. Cunha}
\email{marcony.cunha@uece.br}
\affiliation{Centro de Ciências e Tecnologia, Universidade Estadual do Ceará, 60714-903, Fortaleza, Ceará, Brazil}
\author{R. R. Landim}
\email{renan@fisica.ufc.br}
\affiliation{Departamento de Física, Universidade Federal do Ceará, Caixa Postal 6030, Campus do Pici, 60455-760 Fortaleza, Ceará, Brazil}
\begin{abstract}
In this work, we obtain analytical solutions for a $(3+1)$-dimensional black string and a $(2+1)$-dimensional black hole, both sourced by the Dekel-Zhao dark matter (DM) density profile. Our results indicate that the event horizon radius is sensitive to the inner slope parameter $a$; specifically, beyond a critical threshold, the horizon vanishes, leading to the formation of naked singularities. We find that the DM environment induces curvature singularities in the Ricci and Kretschmann scalars, which are absent in the vacuum BTZ case. Furthermore, an analysis of the effective energy-momentum tensor shows that while the null, weak, and strong energy conditions are strictly satisfied, the dominant energy condition is violated in the lower-dimensional scenario due to the high tangential pressure gradient. We also observe that DM modifies the Hawking temperature and free energy without compromising local or global stability. Notably, the DM distribution transforms the originally constant-curvature BTZ spacetime into a singular one, suggesting that a inherent stiffness of the DM profile is a determinant factor in the causal structure of these solutions.

\noindent{Key words: Black string; Lower-dimensional black hole; Dekel-Zhao, Dark matter.}
\end{abstract}
%
%\pacs{72.80.Le, 72.15.Nj, 11.30.Rd}
%
\maketitle
\section{Introduction}

The enigmatic nature of dark matter (DM), which comprises approximately 29.6\% of the universe's energy-matter content \cite{Planck:2018vyg}, is strongly supported by astrophysical observations such as galaxy rotation curves \cite{Zwicky:1933gu,Rubin:1970zza,Persic:1995ru,Bertone:2016nfn} and its fundamental role in structure formation \cite{Trujillo-Gomez:2010jbn}. Proposed candidates span mass scales from $10^{-22}$ eV fields to primordial black holes \cite{Randall2018DarkMatter}, including WIMPs and axions potentially detectable via solar signatures or satellite orbits \cite{Arguelles:2019ouk,Marsh:2024ury,Tsai:2021lly,Souza:2024ltj}. Furthermore, DM has been investigated as a possible source for wormholes and black holes in both Einstein and modified gravity \cite{Xu:2020wfm, Muniz2022, deSSilva:2024gdc, Carvalho:2022ywl, Mustafa:2023kqt, Zhang:2024zrh, Errehymy:2024lhl, Maurya:2024jos, Hassan:2024xyx}. This quest to identify its nature has led to diverse phenomenological models and density profiles to describe its distribution across various theoretical and observational contexts \cite{Burkert, Navarro, Navarro2, Cintio, Nandi:2004ku, Xu, Mustafa, Haxi}.

The generalized Zhao density profile \cite{Zhao1996, Zhao1997} provides a versatile double power-law framework for modeling dark matter halos, defined as:
\beq \label{density}
\rho(r) = \frac{\rho_{ch}}{\left( \dfrac{r}{r_c} \right)^{\!\!a} \left[1 + \left( \dfrac{r}{r_c} \right)^{\!\!\tfrac{1}{b}} \right]^{b(g - a)}} ,
\eeq
where $\rho_{ch}$ and $r_c$ are the characteristic density and scale radius, while $a$, $b$, and $g$ represent the inner slope, transition sharpness, and outer slope, respectively \cite{Zhao1997}. For configurations where $b=n$ and $g=3+k/n$ ($k, n \in \mathbb{N}$), the gravitational potential, enclosed mass, and velocity dispersion can be derived analytically using elementary functions. Specifically, \cite{Dekel_2017} demonstrated that the configuration $b=2$ and $g=3.5$---the Dekel-Zhao (DZ) profile---provides an exceptional fit across varied halo masses, capturing the cusp-to-core transition more effectively than the Einasto model. The analytical tractability of the DZ profile, further refined for feedback-driven transformations by \cite{Freundlich_2019}, is supported by the seminal work of \cite{Zhao1996, Zhao1997} and the comprehensive framework of \cite{An_2012}, which yields closed-form expressions for gravitational properties in terms of incomplete beta and Fox $H$ functions \cite{Montenegro2012}.

In another direction, cylindrical symmetry has maintained a pivotal role in General Relativity since its inception, spanning from the foundational static and rotating solutions of Levi-Civita, Weyl, Chazy, Curzon, and Lewis \cite{Levi_1917,Weyl_1917,Chazy_1924,Curzon_1924,Lewis_1932} to the spacetimes of cosmic strings, G\"{o}del,  Krasinsky \cite{Vilenkin_1984,Gott_1991,Godel_1949,Krasinski}. More recently, $D$-dimensional black strings have emerged as vacuum generalizations of black holes with translation symmetry, for asymptotically flat or $AdS$ backgrounds \cite{duff1988,emparanPRL,ijmpa2011}. Despite the constraints of the hoop conjecture \cite{Thorne}, the discovery of BTZ and $4$-D black strings \cite{BTZ,Lemos1,Lemos2} facilitated rigorous definitions for mass and angular momentum, sparking extensive investigations across massive, mimetic, non-local gravities, as well as black bounces, noncommutative and optical appearance \cite{Tannukij:2017jtn, Hendi2021, Sheykhi:2020fqf, Singh2018, Pino2019, Ali2020, Lima:2022pvc, Lima:2023arg, Lima:2023jtl, Crispim:2024yjz, Alencar:2025yyl, Furtado:2022tnb}. These objects are not merely theoretical curiosities; their existence is closely associated with the filamentous ``cosmic web'' and dark matter (DM) distributions \cite{Hong2021,Donnan2022}, providing sufficient motivation to explore how DM-sourced black strings—particularly those involving a negative cosmological constant in the dark sector \cite{Calderon2021,Cunha:2022kep}—influence cosmological structure formation \cite{Eisenstein1997}. 

This article aims to analyze the interplay between these relativistic objects and the Dekel-Zhao DM profile, a versatile framework that captures the cusp-to-core transitions observed in astrophysical systems \cite{Zhao1996,Zhao1997,Dekel_2017,Ovgun:2025bol,Khatri:2025lph}. By embedding BTZ black holes and black strings within this DM background, we examine modifications to horizon geometry, Hawking temperature, and thermodynamic stability. Furthermore, we investigate the energy conditions to characterize how DM parameters and analytical potential transformations \cite{An_2012,Freundlich_2019} impact the fundamental properties of these gravitational models.

The paper is organized as follows: In Section II, we derive the analytical solution for a black string sourced by the Dekel-Zhao dark matter profile and analyze its geometric properties, followed by a detailed study of its thermodynamic stability and geodesic motion. In Section III, we extend our analysis to $(2+1)$-dimensional gravity by obtaining the modified BTZ black hole solution. We investigate how the dark matter profile transforms the constant curvature of the vacuum BTZ spacetime into a singular geometry. Furthermore, we perform a rigorous evaluation of the energy-momentum tensor for both models. We demonstrate that while the null, weak, and strong energy conditions are strictly satisfied, the dominant energy condition is violated in the BTZ scenario, suggesting an extreme tangential stiffness inherent to the Dekel-Zhao distribution in lower dimensions. Finally, in Section IV, we present our concluding remarks, discussing the implications of our findings for the cosmic censorship hypothesis and the role of dark matter in non-standard black hole geometries.

%%%
\section{Black String}
Let us consider a line element in (3+1) dimensions of the form  
\beq\label{metric}
ds^2=-f(r)dt^2+\frac{1}{f(r)}dr^2 + r^2 d\phi^2 + \frac{r^2}{\ell^2}dz^2,
\eeq
where $\ell$ is {the fundamental length scale of the model}. %The coordinates are such that $t \in (-\infty,\infty); r \in(-\infty,\infty);\phi \in [0,2\pi]; z \in(-\infty,\infty)$. 
For the black string, we start from the Einstein-Hilbert action, in Planck units, with the addition of the cosmological term, that is,  
\beq
S=\frac{1}{2\kappa}\int d^4x \sqrt{-g}(R-2\Lambda),
\eeq
where $\kappa=8\pi$, $\Lambda=-3/\ell^2$. The Einstein’s equations with the cosmological constant is
\beq \label{EE}
G^{\mu} _{~\nu} + g^{\mu} _{~\nu}\Lambda = \kappa T^{\mu} _{~\nu},
\eeq
which will be solved using the Dekel-Zhao density profile \eqref{density}.

\begin{figure}[h]
    \centering
\includegraphics[width=0.6\textwidth]{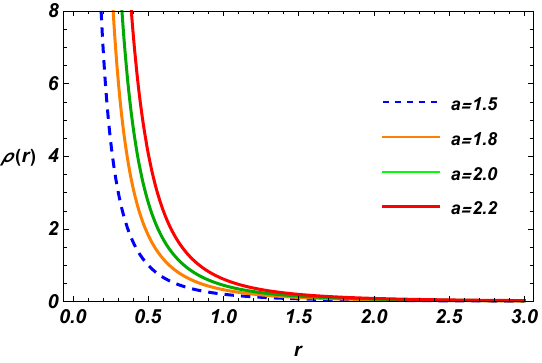}
    \caption{DZ dark matter density profile for some values of $a$. We fixed  $\rho_{ch}=6/5$, $g=7/2$, $b=2$.}
    \label{graficorho}
\end{figure}

This profile is highly effective for modeling dark matter halos; following \cite{Dekel_2017}, we fix the parameters $b = 2$ and $g =\setemeio$, as this specific configuration provides an exceptional phenomenological fit across a wide range of astrophysical systems. In Fig. (\ref{graficorho}), we represent the density profile for some values of $a$.  

Using the conservation of the energy--momentum tensor $\nabla_\mu T^\mu_{~\nu}=0$, we can obtain the components
\begin{eqnarray}
 T^0_{~0} &=&T^1_{~1} = -\rho(r)\label{T00T11} \\ 
 T^2_{~2} & =&T^3_{~3}=-\rho(r)-\frac{r}{2}\frac{d \rho(r)}{dr} \label{T22T33}
\end{eqnarray}
Then, we can write 
\beq T^\mu_{~\mu} = (-\rho,~p_r,~p_L,~p_L). \label{Tmm}\eeq 
In view of Eqs. \eqref{T00T11} and \eqref{T22T33}, we can write $p_r = -\rho$ and 
\begin{equation}\label{plateral}
p_L=\frac{1}{2} \rho_{ch} \left(\frac{r}{r_{c}}\right)^{\!\!-a} \left[1+\left(\frac{r}{r_{c}}\right)^{\!\!1/b}\right]^{b(a -g)-1} \left[\left(\frac{r}{r_{c}}\right)^{\!\!1/b}(g-2) +a -2\right].
\end{equation}
\begin{figure}
%    \centering
    \includegraphics[width=0.60\linewidth]{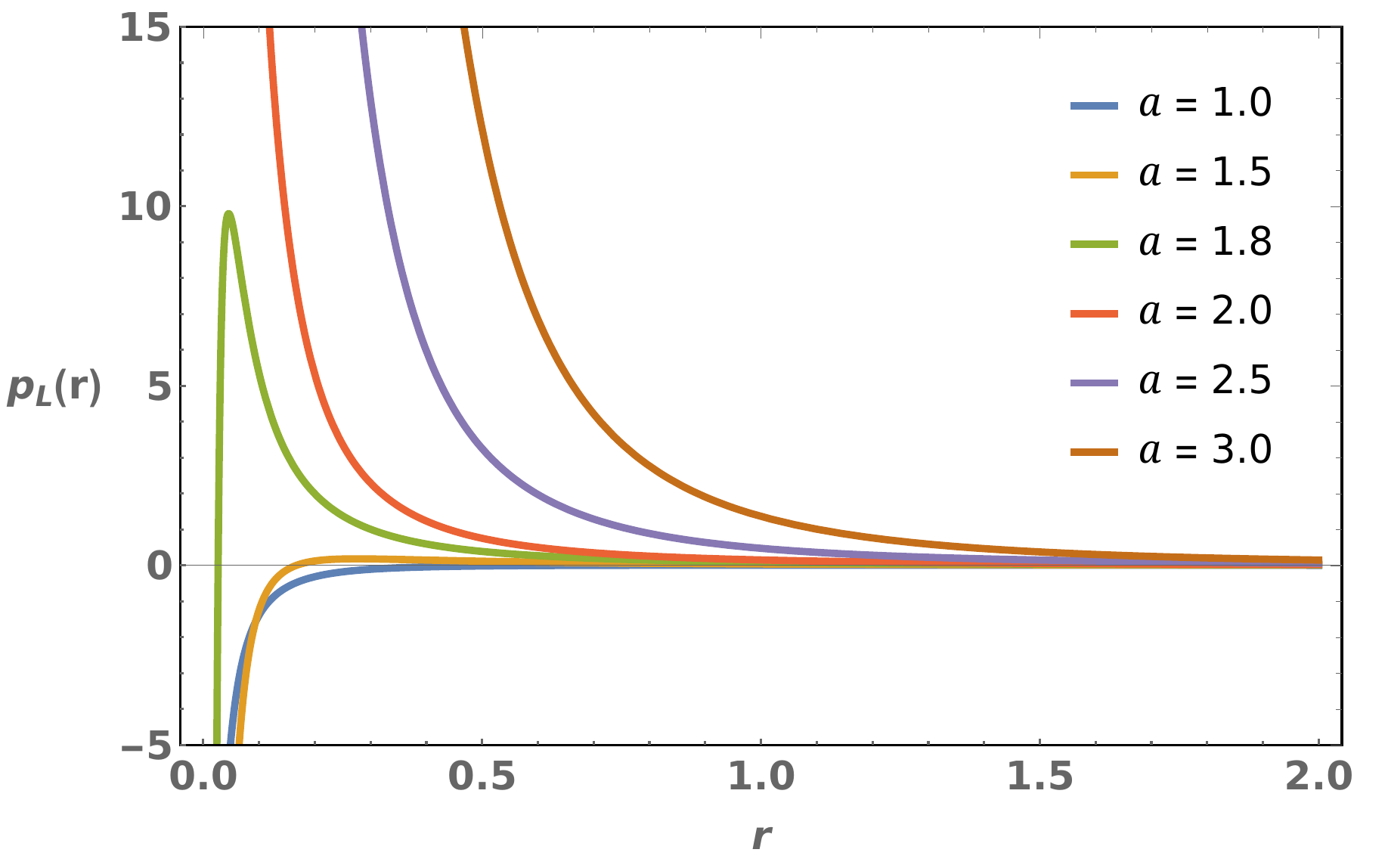}
    \caption{Lateral pressure $p_L$ for several values of $a$. We fixed $\mu=1$, $\ell=1/2$, $\rho_{ch}=6/5$, $g=7/2$, $b=2$, $r_c=3/2$ and $\kappa=8\pi$.}
    \label{figpL}
\end{figure}

\noindent As we can see in Fig. \ref{figpL}, $p_L$ can assume negative values only for $a \in [0,2)$. 

On the other hand, for the metric given by Eq.\eqref{metric}, the $00$ and $11$ components of the Einstein tensor are
\begin{equation} \label{EFE}
G^{0} _{~0} = G^{1} _{~1}= \frac{1}{r^2} \frac{d (r f)}{dr}.
\end{equation}
Therefore, we have
\begin{equation}
  \frac{1}{r^2} \frac{d (r f)}{dr}+\Lambda =-\frac{\kappa \, \rho_{ch}}{\left(\dfrac{r}{r_c} \right)^{\!\!a} \left[1 + \left( \dfrac{r}{r_c} \right)^{\tfrac{1}{b}} \right]^{b(g - a)}}\label{00}
\end{equation}
%
%\begin{align}
%\frac{1}{2r^2} \frac{d(r^2 f')}{dr}+\Lambda=-&\kappa \frac{1}{2} \rho_{ch} \left(\frac{r}{r_{c}}\right)^{-a} \left[1+\left(\frac{r}{r_{c}}\right)^{1/b}\right]^{a b-b g-1} \left[\left(\frac{r}{r_{c}}\right)^{1/b}(g-2) +a -2 \right]\label{22}
%\end{align}
%
In order to solve the above equation, let us define  
\begin{eqnarray}
    f(r)=-\frac{4\mu\ell}{r}+\frac{r^{2}}{\ell^{2}}+r^{2-a}F(z)  , \quad r=r_{c}(-z)^{b}. \label{fr_ansatz}
\end{eqnarray}
Equation (\ref{00}) can be written as 
\begin{equation}
    z\left(1-z\right)^{b(g-a)}F'(z)+b(3-a)\left(1-z\right)^{b(g-a)}F(z)=r_{c}^{a}b\rho_{ch}\label{00z}
\end{equation}
Multiplying both sides by the integrating factor $z^{b(3-a)}$ results in the expression
\begin{equation}
z^{b(3-a)}F'(z)+z^{b(3-a)-1} b (3-a) F(z)=- \frac{\kappa r_c^a \rho_{ch} b z^{b (3-a)-1}}{\left(1-z\right)^{b(g-a)}} , \label{hypergeomb}
\end{equation}
which is better written as
\begin{equation}
\frac{d}{dz}[z^{b(3-a)}F(z)]=-\kappa r_c^a \rho_{ch} b z^{b (3-a)-1}\left(1-z\right)^{b(a-g)} , \label{hypergeomc}
\end{equation}
Integrating both sides, we obtain
\begin{equation}
z^{b(3-a)}F(z) = -\kappa ~r_c^a ~\rho_{ch}~b \int z^{b(3-a)-1} (1-z)^{b(a-g)}dz \label{beta_incompleta}
\end{equation}
%%%%%%%%%%%%%%%%%%%%%%%%%%%%%%%%%%%%%%%%%%%%%%%%%%%%%%%%%%%%%%%%%%%%%%%%%%%%%%%%%%%%%%%%%%%%%%%%%%
%and 
%\begin{align}
%&z\left(1-z\right)F''(z)+(1+5b-2ab)\left(1-z\right)F'(z)+(2b-ab)(3b-ab)z^{-1}\left(1-z\right)F(z)\nonumber
%\\&=\kappa b^{2}(-r_{c})^{a}\rho_{ch}\left[1-z\right]^{ab-bg}\left[-(g-2)+\frac{a-2}{z}\right].	\label{22z}
%\end{align}
%Now, we replace (\ref{00z}) in (\ref{22z})to get
%\begin{equation}\label{hiper22}
%    z\left(1-z\right)F''(z)+\left[1+b(3-a)-\left(1+b(g-a)+b(3-a)\right)z\right]F'(z)-b^{2}(3-a)(g-a)F(z)=0
%\end{equation}
%This is the Hipergeometric equation. In fact, if we  differentiate (\ref{00z}) respect to $z$, after some manipulations, we arrive at
%\begin{equation}
%z\left(1-z\right)F''(z)+\left[ c_h-\left(1+a_h+b_h \right)z\right] F'(z)-a_h b_h F(z)=0, \label{hypergeom}
%\end{equation}
%which is the same as Eq. (\ref{hiper22}) {\color{red}if we identify  $c_{h}=1+b(3-a)$, $a_{h}+b_{h}=b(g-a)+b(3-a)$, and $a_{h}b_{h}=b^{2}(3-a)(g-a)$, with solution given by $F(z)= \,_2F_1(a_h,b_h;c_h;z)$. It is easy to show that  $b_{h}=b(g-a),a_{h}=b(3-a)$ or $b_{h}=b\left(3-a\right),a_{h}=b(g-a)$. Thus, the final form of the solution for $f(r)$ is
%%%%%%%%%%%%%%%%%%%%%%%%%%%%%%%%%%%%%%%%%%%%%%%%%%%%%%%%%%%%%%%%%%%%%%%%%%%%%%%%%%%%%%%%%%%%%%%%%%
The above integral can be written in terms of the Gauss hypergeometric equation as
\begin{equation}
\int z^{\alpha-1} (1-z)^{\beta-1} \, dz = \frac{z^{\alpha}}{\alpha} \, {}_2F_1(\alpha, 1-\beta; \alpha+1; z),
\end{equation}
where $\alpha=b(3-a)$, and $\beta =b(a-g)+1$. Then, we can written the solution $F(z)$ as
\begin{equation}
F(z) = \frac{\kappa~ r^a_c ~\rho_{ch}}{(a-3)}{}~_2F_1\left(b(3-a), b(g-a); b(3-a)+1; z\right) 
\end{equation}
Finally, using Eq. \eqref{fr_ansatz}, the $f(r)$ solution is 
\begin{equation}
f(r) = \left(\frac{r}{r_c}\right)^{\!\!-a}\frac{\kappa \rho_{ch} r^2}{a-3} ~{}_2F_1\left((3-a) b,b (g-a);(3-a) b+1;-\left(\frac{r}{r_c}\right)^{\!\!1 /b}\right) -\frac{4\mu \ell}{r}+\frac{r^2}{\ell ^2}.
\end{equation}
 \begin{figure}
    \centering
\includegraphics[width=0.49\textwidth]{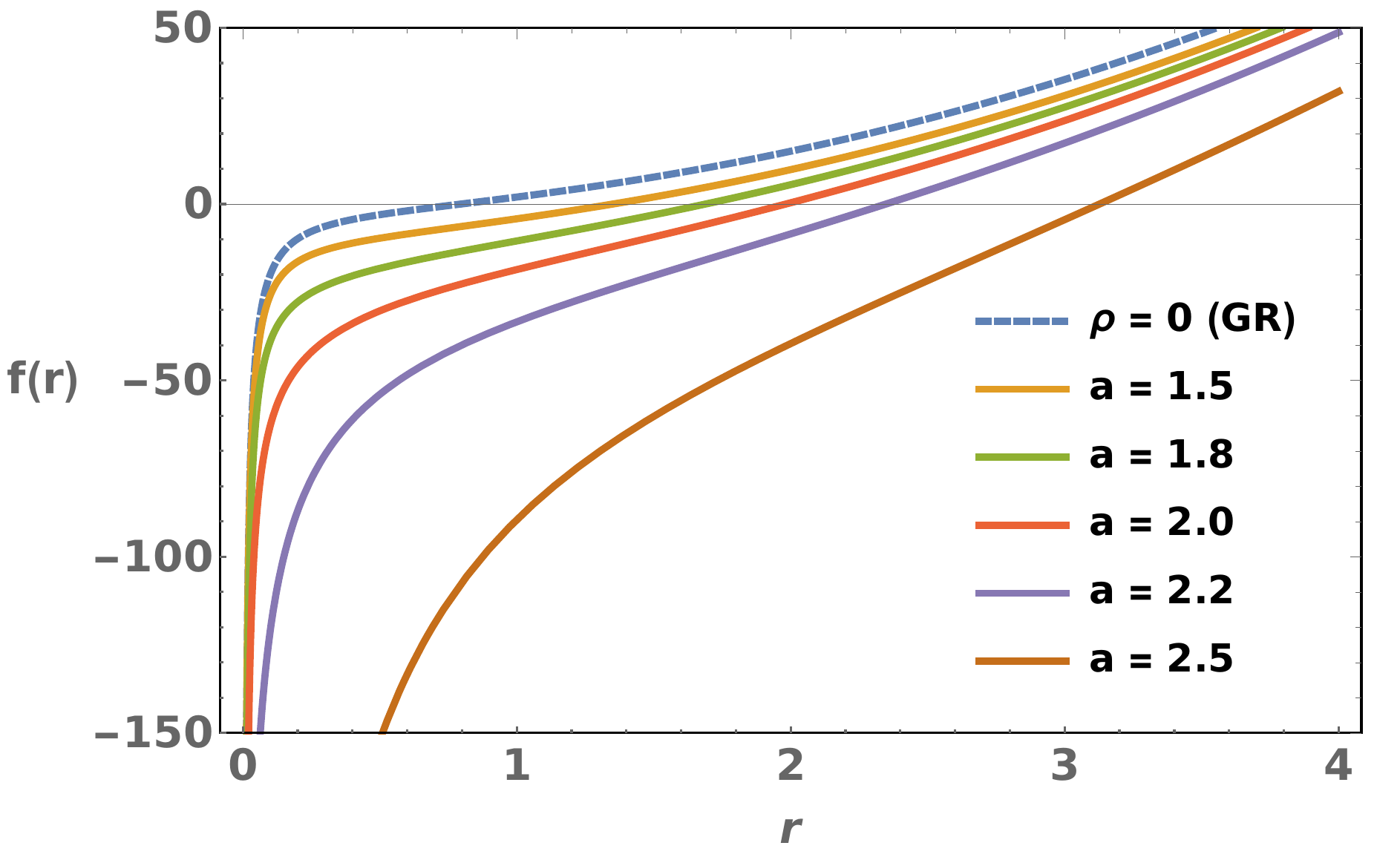}
\includegraphics[width=0.49\textwidth]{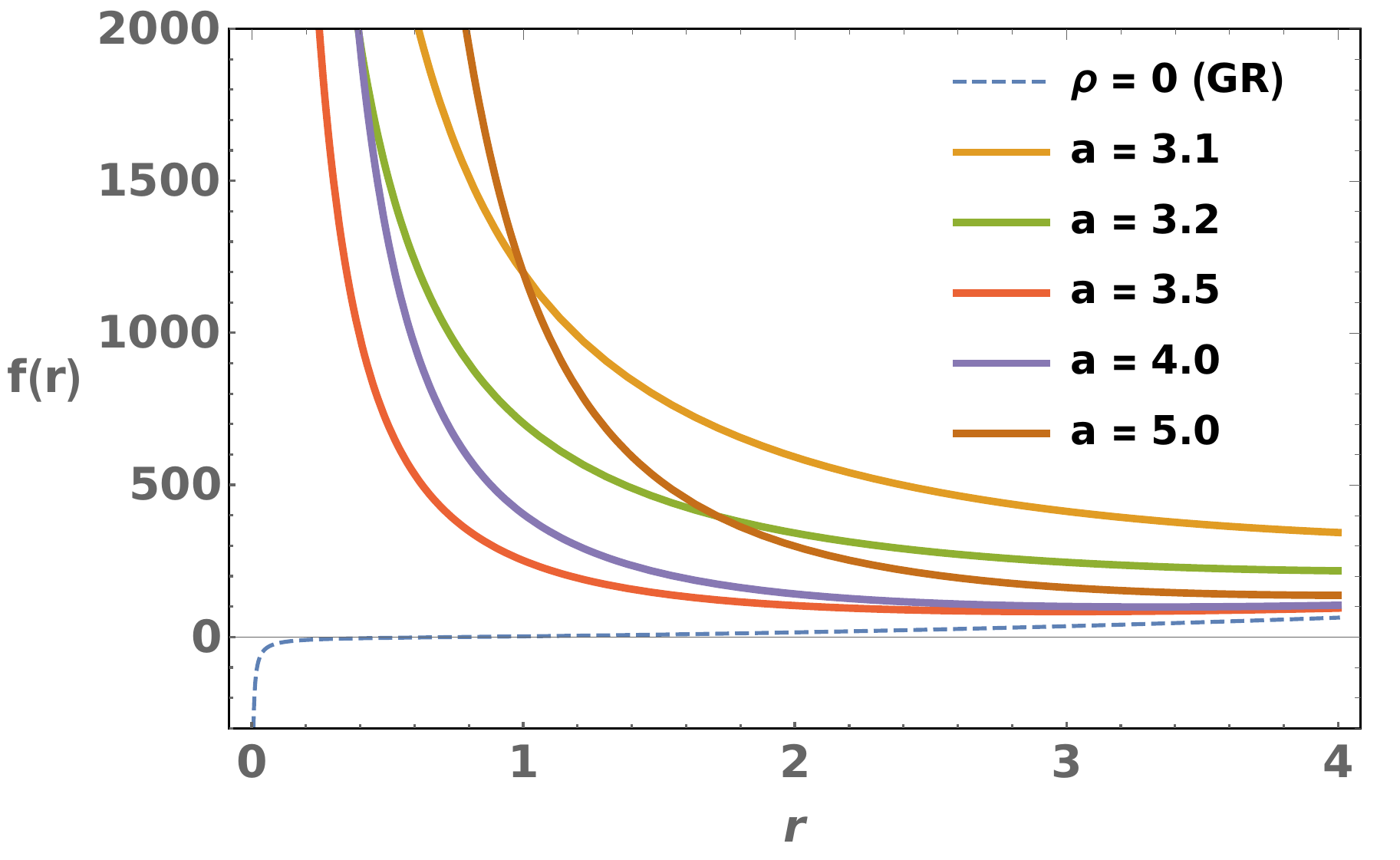}
        \caption{(Left) Metric function $f(r)$ in the presence of DZ dark matter profile for $a<3$. (Right) Metric function for $a>3$ indicating no horizon at all, which can indicate naked singularities. We fixed $\mu=1$, $\ell=1/2$, $\rho_{ch}=6/5$, $g=7/2$, $b=2$, $r_c=3/2$ and $\kappa=8\pi$.}
\label{graficometric}
\end{figure}
For convenience, and without loss of generality, we set the integration constant to zero. Here, $\mu$ represents the linear density of the black string \cite{Cunha:2022kep}. If we set $\rho_{ch}=0$, we recover the vacuum solution, as expected \cite{Lemos1,Lemos2}. The event horizon reaches its maximum value when the parameter $a \rightarrow3$. For $a>3$, there are no horizons at all. Therefore, in order to analyze its thermodynamic properties, we will restrict ourselves to the range $0 < a < 3$.

 In Fig.(\ref{graficometric}), we see that for large values of $r$, $f(r)$ tends to the vacuum solution, and also, the presence of dark matter alters the event horizon radius depending on the parameter $a$. Now for $r/r_c \ll 1$, we have  
\begin{equation}
f(r)\approx -\frac{4\mu \ell}{r}+\frac{r^2}{\ell ^2} + \frac{\kappa \rho_{ch}r_{c}^a}{a-3} r^{2-a} .
\end{equation}
{If $a<2$, the second and third terms tend to zero when $r\rightarrow 0$ and $f(r)\approx -4\mu \ell /r$, which is the vacuum solution near the origin. When $2<a<3$, we can write

\begin{equation}
    f(r)\approx -\frac{4\mu\ell}{r}+\frac{\kappa\rho_{ch}r_c^a}{a-3}\frac{1}{r^{a-2}}\nonumber
\end{equation}

\noindent In this case, the second term goes to infinity but first term goes faster when $r\rightarrow 0$ and we obtain again the vacuum solution. Thus, as shown in Fig. \eqref{graficometric}, taking the asymptotic limits, we recover Lemos’ solution, predicted by General Relativity.}

\subsection{Energy Conditions}
\begin{figure}
    \includegraphics[width=0.6\linewidth]{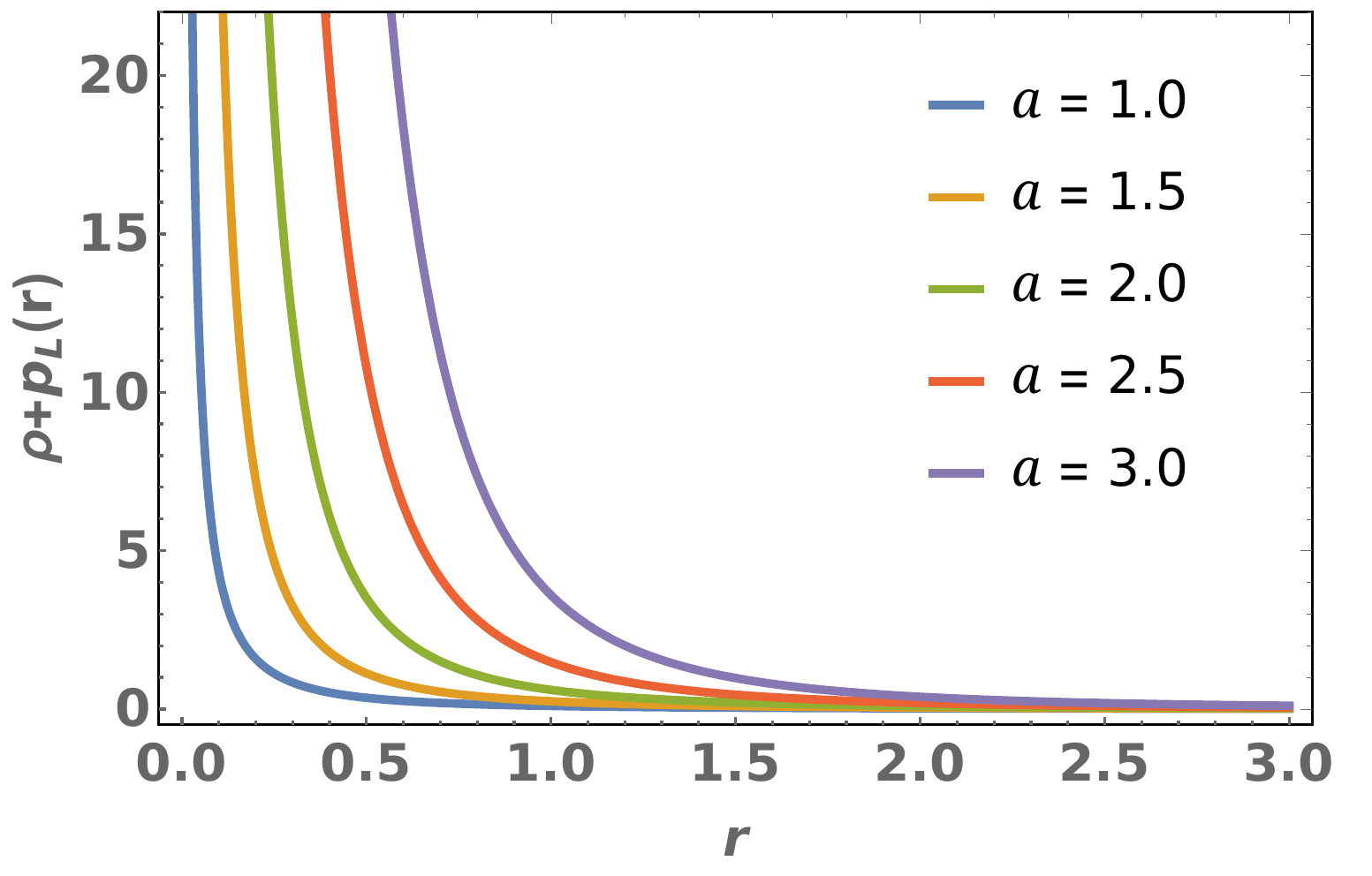}
    % Segunda subfigura
   %\hspace{0.5cm}\begin{subfigure}{0.42\textwidth}
      %  \centering
  %  \includegraphics[width=\linewidth]{graficoDEC.pdf}
     %   \label{graficoDEC}
  %  \end{subfigure}
   % \vspace{-0.88cm}
    \caption{NEC for the parameters $\rho_{ch}=6/5$, $g=7/2$, $b=2$ and $\kappa=8\pi$. We use $a =1, 1.5,2, 2.5, \text{and } 3$. The case $\rho + p_r =0$ (trivially null).}
            \label{NEC}
        \end{figure}

Analyzing Eqs. \eqref{density}, \eqref{Tmm}, and \eqref{plateral} we can discuss the energy conditions.
\begin{enumerate}
\item \textbf{Null Energy Condition (NEC)}: $\rho +p_i\geq0$. \\
As seen previously, $p_r=-\rho$, then $\rho+p_r=0$. For $\rho_{ch} >0 \text{ and }r \geq 0$, it is straightforward to show that $\rho+p_L \geq 0$ $\forall ~a$, $b$, and $g$. Then, NEC is always satisfied [see Fig. \ref{NEC}].

\item \textbf{Weak Energy Condition (WEC)}: $\rho \geq0, \rho +p_i\geq0$. \\
According to Eq. \eqref{density}, whenever $\rho_{ch} > 0$, we have $\rho \geq 0$ as shown if Fig.\eqref{graficorho}. Since the NEC is satisfied, consequently the WEC is also satisfied.

\item \textbf{Strong Energy Condition (SEC)}: $\rho+p_L\geq0, p_L\geq0$. \\
Although $\rho + p_L(r) \geq 0$ for all $r$ in Fig.\eqref{NEC}, Fig.\eqref{figpL} shows that the strong energy condition is violated whenever $a < 2$.

\item \textbf{Dominant Energy Condition (DEC)}: $\rho\geq |p_i|$.\\
The dominant energy condition imposes a stronger constraint by requiring that $\rho \ge |p_i|$. In the present anisotropic configuration, this condition reduces to $\rho \ge |p_L|$. Our numerical analysis shows that, for all values of $r$ and $a$, the lateral pressure remains bounded by the energy density, as illustrated in Fig. \ref{DEC_grad} below. 

\end{enumerate}

\begin{figure}[ht!]
\includegraphics[width=0.6\textwidth]{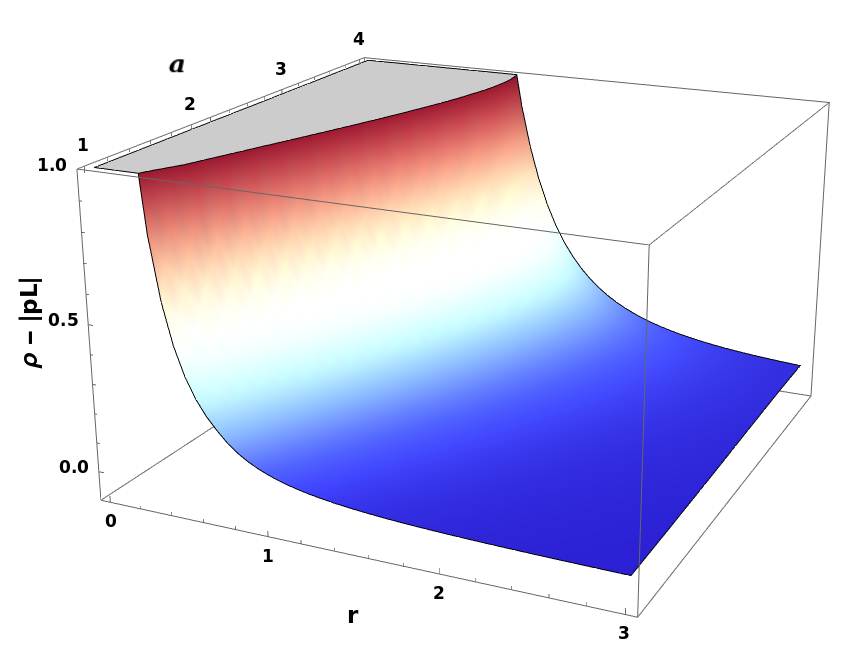}
\caption{Continuous family of curves confirming  DEC, $\rho - |p_L| \ge 0$, as a function of the radial coordinate $r$, for the parameter range $0 \le a \le 4$. We fixed $\mu=1$, $\ell=1/2$, $\rho_{ch}=6/5$, $g=7/2$, $b=2$, $r_c=3/2$ and $\kappa=8\pi$ .}
\label{DEC_grad}
\end{figure}
%%%%
%%%%
%

%\end{enumerate}
%
\subsection{Curvature invariants}
\begin{figure}[h!]
%    \centering
\includegraphics[width=0.6\textwidth]{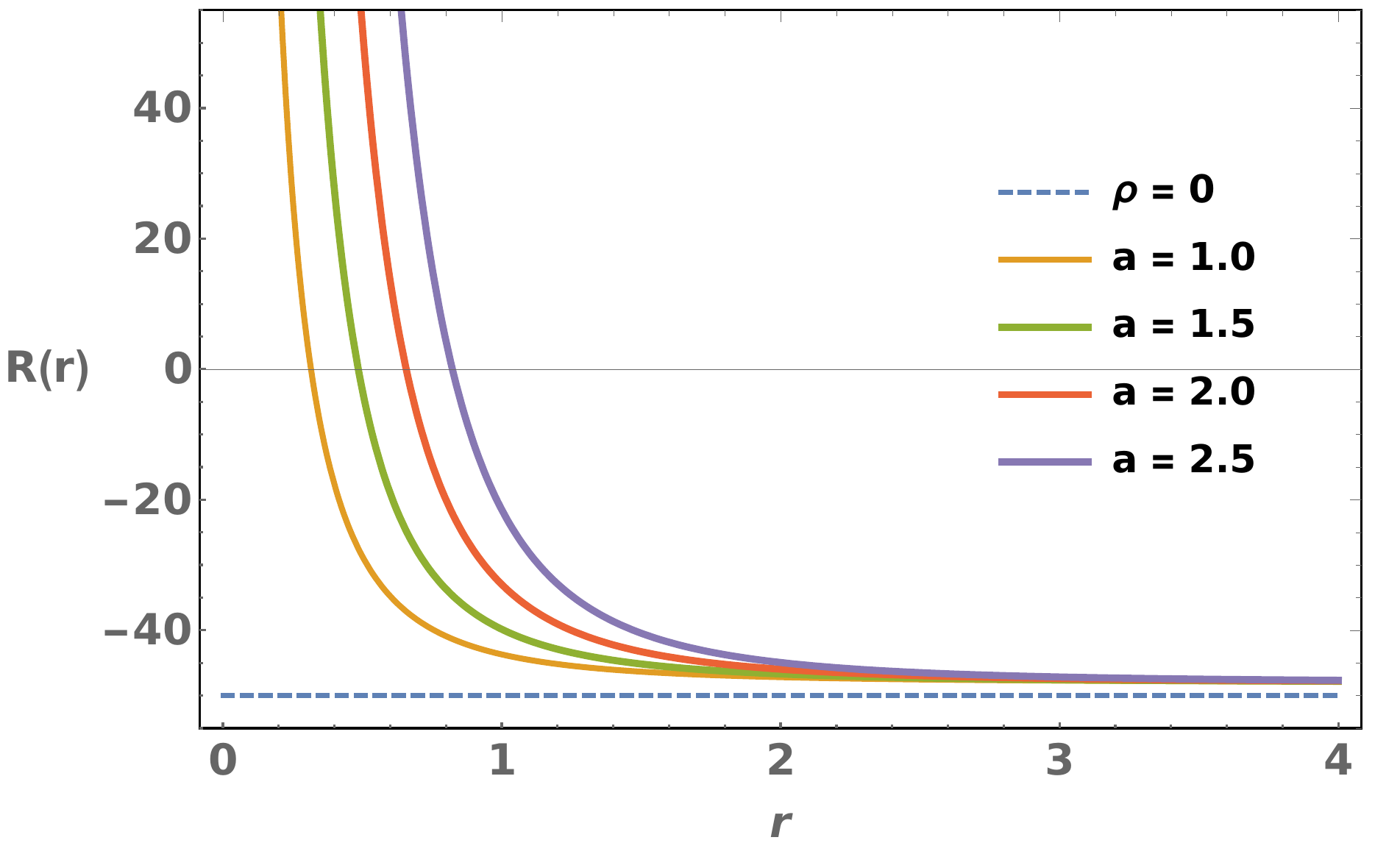}
\caption{Ricci scalar for several values of $a$. We fixed $\mu=1$, $\ell=1/2$, $\rho_{ch}=6/5$, $g=7/2$, $b=2$, $r_c=3/2$ and $\kappa=8\pi$.}\label{graficoRICCI}
\end{figure}

Now, let us analyze the curvature invariants, which will provide us with more precise information about possible singularities, and whether or not they are affected by the presence of DM. First, we have the Ricci scalar,  
\begin{eqnarray}
R(r)=\kappa_{\rho ch}\left(\frac{r}{r_c}\right)^{-a} \left[1+\left(\frac{r}{r_c}\right)^{1/b}\right]^{b(a-g)-1} \left[4-a-(g-4) \left(\frac{r}{r_c}\right)^{1/b}\right]-\frac{12}{\ell ^2}
\end{eqnarray}
where, for $\rho_{ch}=0$, we recover the vacuum solution, $R =-12/\ell^2$, Fig. \eqref{graficoRICCI}.

We see that, when $r \rightarrow 0$, $R(r)$ diverges, which may indicate a singularity there. For $a>0$, the singularity is strengthened, while for $a<0$, as already expected, $R(r)$ approaches the vacuum solution. 
For $r/r_c<<1$, we have
\beq
R(r) \approx \kappa_{\rho ch}\left(\frac{r}{r_c}\right)^{-a}\left[4-a-g\left(\frac{r}{r_c}\right)^{1/b} \right]-\frac{12}{\ell ^2}.
\eeq
Indeed, for $a>0$, the Ricci scalar is singular, while for $a \leq 0$ it is regular. To confirm the existence of a physical singularity, we compute the Kretschmann scalar $K=R_{\mu \nu \sigma \rho}R^{\mu \nu \sigma \rho}$ and the squared Ricci scalar $R^{2}=R_{\mu\nu}R^{\mu\nu}$.

\begin{figure}[ht!]
    \centering
    \begin{subfigure}{0.47\textwidth}
        \centering
        \includegraphics[width=\linewidth]{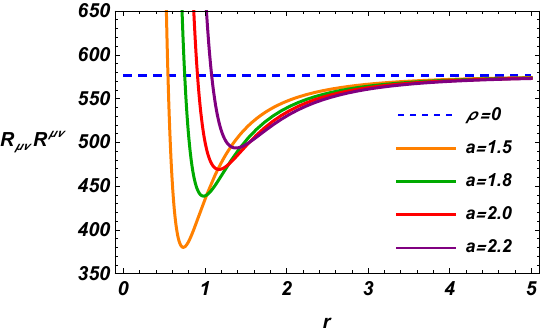}
        \label{graficoRICCI2}
    \end{subfigure}
    \hspace{0.05\textwidth}
    \begin{subfigure}{0.44\textwidth}
        \centering
        \includegraphics[width=\linewidth]{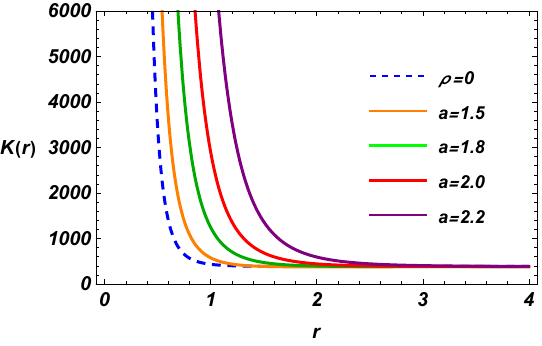}
        \label{graficoK}
    \end{subfigure}
    \caption{$R^{\mu\nu}R_{\mu\nu}$ and $K(r)$, respectively, for some values of $a$. We fixed $\mu=1$, $\ell=1/2$, $\rho_{ch}=6/5$, $g=7/2$, $b=2$, $r_c=3/2$ and $\kappa=8\pi$.}
    \label{KretRicci2}
\end{figure}

The squared Ricci scalar is given by
\begin{align}
%\begin{split}
R_{\mu\nu} R^{\mu\nu} &= \frac{1}{2 \ell^4} \left\lbrace 
4 \left[\kappa \rho_{ch} \ell^2 \left( \frac{r}{r_c} \right)^{\!\!-a}
\left(1 + \left( \frac{r}{r_c} \right)^{\!\!\tfrac{1}{b}} \right)^{\!\!\! b(a - g)} \!\!\! - 3\,\right]^2 
+ \left(\frac{r}{r_c} \right)^{\!\!-2a} \left[1 + \left( \frac{r}{r_c} \right)^{\!\!\frac{1}{b}} \right]^{-2 - 2 b g}\right.\nonumber\\ 
&\times \left[ \kappa \rho_{ch} \ell^2
\left( 1 + \left( \frac{r}{r_c} \right)^{\frac{1}{b}} \right)^{\!\!a b}
\left(a + g \left( \frac{r}{r_c} \right)^{\!\!\tfrac{1}{b}} - 2 \left( 1 + \left( \frac{r}{r_c} \right)^{1/b} \right)
\right)\right]^2\nonumber\\
&+\left. 6 \left( \frac{r}{r_c} \right)^{\!a} \left[1 + \left( \frac{r}{r_c} \right)^{\!\tfrac{1}{b}} \right]^{1 + b g} 
\right\rbrace
%\end{split}
\end{align}
where, as shown in Fig. \eqref{KretRicci2} (left panel), there are singularities that become stronger as $a$ increases.
%\begin{figure}[hb!]
%   \centering
%\includegraphics[width=0.6\textwidth]{graficoRICCI2.pdf}
%        \caption{Squared Ricci scalar scalar for some values of  $a$. We fixed $\mu=1$, $\ell=1/2$, $\rho_{ch}=6/5$, $g=7/2$, $b=2$, and $\kappa=8\pi$.}
%\label{graficoRICCI2}
%\end{figure}
%

For $r/r_c \ll 1$, we have the following expression
\begin{equation}
R^{\mu\nu}R_{\mu\nu} \approx \frac{1}{2\ell^4} \Bigg\lbrace
    4\left[
        -3 + \ell^2\kappa\rho_{ch} \left(\frac{r}{r_c}\right)^{-a}
    \right]^2
    + \left(\frac{r}{r_c}\right)^{-2a} \left[
        6\left(\frac{r}{r_c}\right)^a + \ell^2\kappa\rho_{ch}(a - 2)
    \right]^2
\Bigg\rbrace
\end{equation}
where, for $a=0$, there is no longer a singularity, and the scalar takes a constant value
\beq
R^{\mu\nu}R_{\mu\nu} \approx \frac{1}{2\ell^4} \left[4\left(-3 + \ell^2\kappa\rho_{ch} \right)^2\left(  6 -2\ell^2\kappa\rho_{ch}\right)^2
\right].
\eeq
Thus, values of $a$ greater than $2$ strengthen the singularity in the squared Ricci scalar for small values of $r$.

Now we show the Kretschmann scalar Fig.\eqref{KretRicci2} (right panel), which will give us more precise information about the singularities. It is given by
{\small
\begin{eqnarray}
K(r)&=&\frac{\left({r}/{r_c}\right)^{\!-2 a}}{(a-3)^2 r^6 \ell^4}\Bigg\lbrace 4\Bigg[ (a-3) \left(4 \left(\frac{r}{r_c}\right)^{\!a} \ell^3 \mu +  r^3 \left(2\left(\frac{r}{r_c}\right)^a - \left(1 + \left(\frac{r}{rc}\right)^{\frac{1}{b}}\right)^{\! b (a - g)}\ell^2 \kappa \rho_{ch}\right) \right)\nonumber\\
&-&r^3\ell^2\kappa \rho_{ch}~{}_2F_1\Bigg]^2 \!\!  + 4 \Bigg[ (a-3) \left(\frac{r}{r_c}\right)^{\!a}\!\!(r^3-4\mu \ell^3)+ r^3\ell^2\kappa\rho_{ch}~{}_2F_1 \Bigg]^{\!2} \!\! +\bigg(\!1+\left(\frac{r}{r_c}\right)^{\!\!\tfrac{1}{b}}\bigg)^{2b (a-g)-2}\nonumber\\
&\times&\Bigg[(a-3)\Bigg(\!\!-8\mu\ell^3\left(\frac{r}{r_c}\right)^{\!a} \left(1+\left(\frac{r}{r_c} \right)^{\!\!\frac{1}{b}} \right)^{\!\! 1+b(g-a)}\!\!\!\!\! + ~~r^3\Bigg( 2\left(\frac{r}{r_c} \right)^{\!a}\bigg( 1+\left(\frac{r}{r_c}\right)^{\!\tfrac{1}{b}} \bigg)^{1+b(g-a)} \nonumber\\
&+&\left. \ell^2\kappa\rho_{ch} \right.\bigg(a+g\bigg(\frac{r}{r_c}\bigg)^{\!\!\tfrac{1}{b}}\bigg)\Bigg)\Bigg) + 2r^3\ell^2 \kappa\rho_{ch}\left(1+\left(\frac{r}{r_c}\right)^{\!\frac{1}{b}} \right)^{\!\! 1+b(g-a)} {}~_2F_1 \Bigg]^2 \Bigg\rbrace
\end{eqnarray}
}
%%%%%%%%%%%%%%%%%%%%%%%%%%%%%%%%%%%%%%%%%%%%%%%%%%% FIM DA EQUAÇÃO
\noindent where ${}_2F_1 \equiv {}_2F_1\Big[(3-a)b, b(g-a); 1+(3-a)b; -\left(\dfrac{r}{r_c}\right)^{\!\!\frac{1}{b}}\Big]$ and the $r^{-6}$ singularity is recovered whenever $\rho_{ch} = 0$. 

For the small $r/r_c\ll 1$, we have
\begin{eqnarray} \label{approxkret}
K(r) \approx \frac{1}{(a - 3)^2 \ell^4 \, r^6 } \left\lbrace \left[8\ell^3(a-3)\right]^2+ 4\left[1+4(a-3)\ell^3 \right]^2 + \left[1 + 8 (a - 3) \ell^3 \right]^2 \right \rbrace.
\end{eqnarray}
{As usual, we observe the physical singularity at $1/r^{6}$, which becomes stronger for $a \approx 3$.} Thus, the presence of dark matter enhances the singularity in the Kretschmann scalar.
%\begin{figure}[ht!]
%    \centering
%\includegraphics[width=0.6\textwidth]{graficoK.pdf}
%\caption{Kretschmann scalar for some values of  $a$. We fixed $\mu=1$, $\ell=1/2$, $\rho_{ch}=6/5$, $g=7/2$, $b=2$, and $\kappa=8\pi$.}
%\label{graficoK}
%\end{figure}

In summary, our analysis demonstrates that both the Ricci scalar and the squared Ricci scalar remain regular for $a=0$, indicating the absence of curvature singularities in this case. However, for $a>0$, their singular behavior becomes increasingly pronounced, showing that the introduction of dark matter leads to stronger divergences in these curvature invariants. On the other hand, the Kretschmann scalar is singular in all cases, and its divergence also grows with increasing $a$. This indicates that, while the intrinsic singularity of the spacetime, the presence of dark matter not only maintains the singularity but also amplifies its intensity, reflecting a deeper influence of dark matter on the extreme curvature of the spacetime near the origin.

\subsection{Thermodynamics}  
In this section, we analyze the Hawking temperature, the entropy per unit length, the heat capacity per unit length, and the Helmholtz free energy per unit length. First, the Hawking temperature is given by  
\beq \label{thawking}
T_H=\frac{\kappa}{2 \pi}=\frac{1}{4\pi}\left. \frac{df(r)}{dr}\right|_{r = r_h},
\eeq  
where $r_h$ is the event horizon position, obtained from  
\begin{eqnarray}
\left(\frac{r_h}{r_c}\right)^{\!\!-a}\frac{\kappa  \rho_{ch} {r_h}^2}{a-3} \, _2F_1\!\Big[\!(3-a) b,b (g-a);(3-a) b+1;-\left(\frac{r_h}{r_c}\right)^{\!1/b}\Big] -\frac{4\mu \ell}{r_h}+\frac{{r_h}^2}{\ell ^2} =0,
\end{eqnarray}  
However, an analytical solution for $r_h$ is not possible.  

\begin{figure}[ht!]
    \centering
\includegraphics[width=0.6\textwidth]{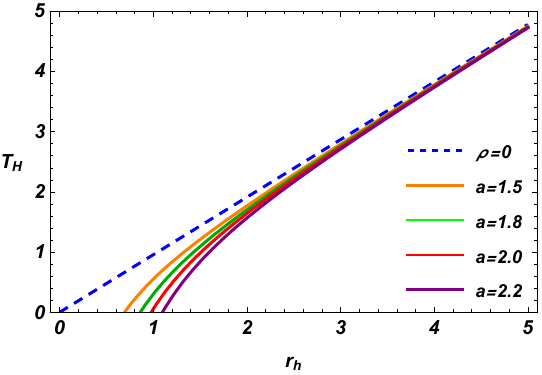}
        \caption{Hawking temperature for some values of $a$. We fixed $ \ell=1/2, \rho_{ch}=6/5, g=7/2, b=2   \ \text{and} \ r_c=3/2.$}
\label{graficoTH}
\end{figure}

Thus, we obtain the Hawking temperature for the black string surrounded by dark matter, 
\beq
T_H=\frac{1}{4\pi} \Bigg \lbrace \frac{3r_h}{\ell ^2} -r_h \left[ 1 + \left(\frac{r_h}{r_c}\right)^{1/b}\right]^{b(a-g)}\left(\frac{r_h}{r_c}\right)^{-a}\kappa \rho_{ch} \Bigg\rbrace.
\eeq  
The following plot, Fig.\eqref{graficoTH}, shows the Hawking temperature as a function of the event horizon.  

We also note that for $\rho_{ch}=0$, we recover the vacuum black string solution. The temperature is always increasing, and for large values of $r_h$, it approaches the vacuum temperature $3r_h/4\pi \ell^2$. When the temperature reaches its minimum in vacuum, the black string would completely evaporate. However, in the presence of dark matter, there is a remnant mass density, as seen in Fig.(\ref{graficoTH}).  

We can directly compute the entropy from the expression $ds=d\mu /T_H$, yielding  
\beq
s=\frac{\pi r_h^{2}}{2\ell}.
\eeq  
This is the usual result for vacuum black strings, indicating that the presence of dark matter does not affect the entropy of the black string, since it depends only on the geometry of the horizon and is proportional to its area.\cite{Bekenstein:1973ur}.

The heat capacity is computed from $c=d\mu/dT_H$, giving  
\beq
c=\frac{
    \pi r_h^2 \left[ 1 + \left( \frac{r_h}{r_c} \right)^{\tfrac{1}{b}} \right]
   \left\lbrace
        3 \left(\frac{r_h}{r_c} \right)^a
        - \left[ 1 + \left( \frac{r_h}{r_c} \right)^{1/b} \right]^{b(a - g)}
        \ell^2 \kappa \rho_{\text{ch}}
   \right\rbrace
}{
    3 \left(\frac{r_h}{r_c}\right)^{\!a}
    \left[ 1 + \left(\frac{r_h}{r_c}\right)^{\tfrac{1}{b}}\right]\ell
    +
    \left[
        a + (g - 1) \left(\frac{r_h}{r_c}\right)^{\tfrac{1}{b}}-1
    \right]
 \left[ 1 + \left(\frac{r_h}{r_c}\right)^{\tfrac{1}{b}} \right]^{b(a - g)}\!\!\! \ell^3 \kappa \rho_{ch}},
\eeq  
which for $\rho_{ch}=0$ reduces to the usual solution $c=\pi r_h ^{2} / \ell $. Regions with $c<0$ indicate thermodynamic instability, while regions with $c>0$ indicate local thermodynamic stability. Divergences would signal a phase transition.  

\begin{figure}[ht!]
    \centering
\includegraphics[width=0.6\textwidth]{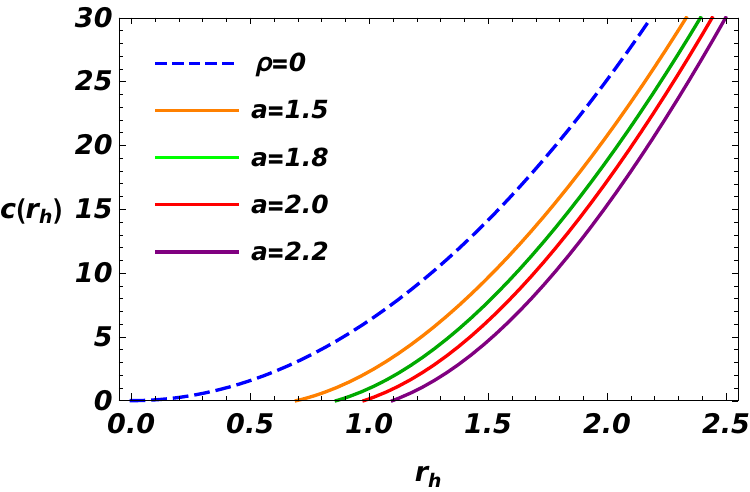}
        \caption{Heat capacity for some values of $a$. We fixed $\mu=1, \ell=1/2, \rho_{ch}=6/5, g=7/2, b=2, r_c=3/2 \ \text{and} \ \kappa=8\pi$.}
\label{graficoHC}
\end{figure}  

In Fig.(\ref{graficoHC}), we observe that the heat capacity is always positive for all analyzed values of $a$, indicating local thermodynamic stability, and there is no phase transition. Thus, although dark matter affects the temperature and heat capacity, the stability remains unaffected, unlike other density profiles \cite{Cunha:2022kep}.  

\begin{figure}[ht!]
    \centering
\includegraphics[width=0.6\textwidth]{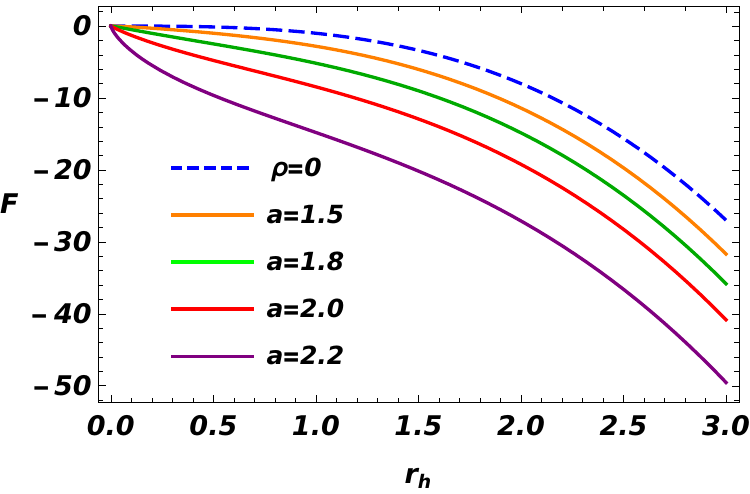}
        \caption{Free energy for some values of $a$. We fixed $\mu=1, \ell=1/2, \rho_{ch}=6/5, g=7/2, b=2, r_c=3/2 \ \text{and} \ \kappa=8\pi$.}
\label{graficoFE}
\end{figure}

Now, calculating the Helmholtz free energy per unit length, $F=\mu - T_{H} s$, we observe that the presence of dark matter modifies the free energy. 
\bea
F &=&- \frac{r_h ^3}{8 (a-3) \ell^3}\left(\frac{r_h}{r_c}\right)^{-a}\Biggr\lbrace \Bigg[\left(\frac{r_h}{r_c}\right)^{a} - \Bigg(1+\left(\frac{r_h}{r_c}\right)^{\tfrac{1}{b}}\Bigg)^{b(a-g)} \ell^2 \kappa \rho_{ch}\Bigg](a-3) \nonumber\\
&-&2\ell^2 \kappa \rho_{ch}\, _2F_1\!\Bigg(\!(3-a) b, b (g-a);(3-a) b+1;-\left(\frac{r_h}{r_c}\right)^{\tfrac{1}{b}}\Bigg) \Bigg\rbrace
\eea
As shown in Fig.\eqref{graficoFE}, $F<0$, and increasing the parameter $a$ decreases the value of $F$. This behavior indicates that there is no phase transition in the black string for this dark matter density and also suggests global stability. Again, for large values of $r_h$, the free energy tends to the vacuum solution.  

\subsection{Geodesics}  
Now we study the trajectories of particles in this spacetime. To do so, we analyze the geodesics. From the metric \eqref{metric}, we can write the Lagrangian  
\beq
\mathcal{L}=g_{\mu\nu}\dot{x^\mu}\dot{x^\nu}=-f(r) \dot{t}^2+\frac{1}{f(r)}\dot{r}^2+r^2\dot{\phi}^2+\frac{r^2}{\ell^2}\dot{z}^2,
\eeq  
where the dot represents derivative with respect to the affine parameter. Using the Euler-Lagrange equations, we obtain the geodesic equations for massless particles:  
\beq \label{geodesics}
\begin{split}
&\ddot{t} +\frac{f'(r)}{f(r)}\dot{r}\dot{t}=0,\\
&\ddot{r}+\frac{1}{2}f(r)f'(r)\dot{t}^2 -\frac{f'(r)}{2f(r)}\dot{r}^2-r\left[f(r)\dot{\phi}^2-\frac{1}{\ell^2} f(r)\dot{z}^2\right]=0, \\
&\ddot{\phi}+\frac{2}{r}\dot{r}\dot{\phi}=0, \\
&\ddot{z}+\frac{2}{r}\dot{z}\dot{r}=0.
\end{split}
\eeq  

From these equations, we can extract the following conserved quantities:  
\beq
\begin{split}
\dot{t} f(r)=E, \\
\dot{\phi} r^2=L,  \\
\frac{\dot{z} r^2}{\ell^2}  =p ,
\end{split}
\eeq  
where $E$ is the total energy of the particle, $L$ is the angular momentum along the $z$-axis, and $p$ is the momentum along the $z$-axis. Using these relations, the radial geodesic can be written as  
\beq
\dot{r}^2+V_{eff}=E^2 \quad ,\quad V_{eff}=f(r)\left( \frac{L^2}{r^2}+\frac{p^2\ell^2}{r^2}-\epsilon\right),
\eeq  
with $\epsilon=\mathcal{L}=0,-1$ for massless and massive particles, respectively.  

\begin{figure}[ht!]
    \centering
\includegraphics[width=0.6\textwidth]{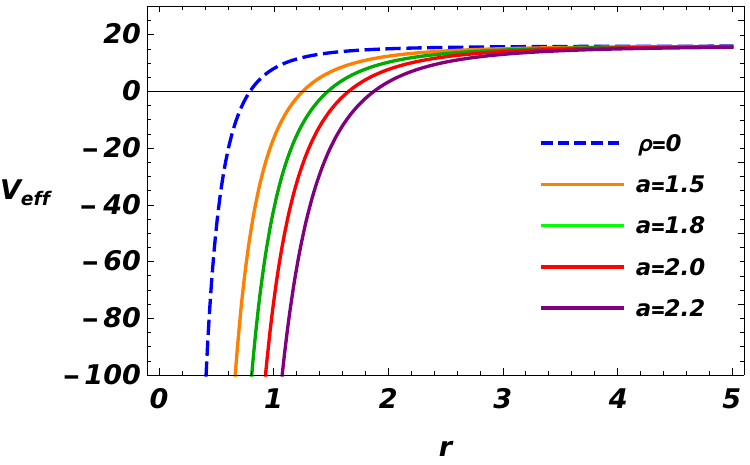}
        \caption{Effective potential for massless particles for some values of $a$. We fixed $\mu=1, L=1, \ell=1/2, \rho_{ch}=6/5, g=7/2, b=2, r_c=3/2 \ \text{and} \  \kappa=8\pi$.}
\label{graficoG0}
\end{figure}

For circular photon orbits in the constant $z$-plane
$(\epsilon=0, \dot{z}=0)$, we impose the condition $V'_{eff}(r_0)=0$ with $\dot{r}_0=0$. The effective potential and its derivative are given by
{\small
\beq
V_{eff}(r)=\left(\frac{r}{r_c}\right)^{\!\!-a}\!\frac{L^2 \kappa  \rho_{ch}}{3-a} \, _2F_1\!\left(\!(3-a) b,b (g-a);(3-a) b+1;-\left(\frac{r}{r_c}\right)^{\tfrac{1}{b}}\right)
+\frac{L^2}{\ell^2}-\frac{4\mu\ell L^2}{r^3}, 
\eeq  
\bea
V'_{eff}(r)&=&\frac{12L^2\ell \mu}{r^4}-\frac{L^2 \kappa \rho_{ch}}{r}\left(\frac{r}{r_c}\right)^{-a}\left[1+\left(\frac{r}{r_c}\right)^{\tfrac{1}{b}}\right]\hspace{7.1cm} \nonumber\\
&+&\left(\frac{r}{r_c}\right)^{-a}\frac{3L^2 \kappa \rho_{ch}}{r(3-a)} \, _2F_1\left((3-a) b,b (g-a);(3-a) b+1;-\left(\frac{r}{r_c}\right)^{\tfrac{1}{b}}\right).
\eea
}
We observe that there is no $r_0$ such that $V'_{eff}(r_0)=0$. This implies that no photon sphere exists, i.e., there are no circular orbits for the black string generated by dark matter. % maintaining the usual result for black strings \cite{Darla2023Rainbow,Lima:2023arg,Lima:2023jtl}.
Figure \eqref{graficoG0} shows the behavior of the effective potential for some values of $a$.  

For small values of $r$, we see that in the presence of dark matter, the effective potential tends to decrease.

\section{Lower-Dimensional Black Hole}  
Now, we analyze a lower-dimensional BTZ-like spacetime generated by the DZ dark matter profile \cite{Zhao1996,Zhao1997}. This geometry describes a $(2+1)$-dimensional black hole and is characterized by the metric \cite{BTZ}:  
\beq \label{BTZeq}
ds^2=-f(r)dt^2+\frac{1}{f(r)}dr^2+r^2 d\phi ^2.
\eeq  

The components of the Einstein tensor are  
\beq
G^0_0=G^1_1=\frac{f'(r)}{2r} \quad , \quad G^2_2= \frac{f''(r)}{2}.
\eeq  

To solve the $00$ component of Einstein's equation, we define  
\beq\label{ansatzBTZ}
f(r)=-M + \frac{r^2}{\ell^2} + r^{2-a} F(z), \quad r=r_c(-z)^b,
\eeq  
where $M$ is the black hole mass. This leads to  
\beq
F'(z)+(2-a)\frac{b}{z}F(z)=-2\kappa \frac{ r_c^a \rho_{ch} b}{z(1-z)^{b(g-a)}}.
\eeq  
Multiplying both sides by $z^{b(2-a)}$ we obtain the following expression
\beq
z^{b(2-a)}F'(z)+z^{b(2-a)-1}b(2-a) F(z)=-2\kappa \frac{ r_c^a \rho_{ch} b z^{b(2-a)-1}}{(1-z)^{b(g-a)}},
\eeq
which can be written as
\beq
\frac{d}{dz}[z^{b(2-a)}F(z)]=-2\kappa r_c^a \rho_{ch} b z^{b (2-a)-1}\left(1-z\right)^{b(a-g)}.
\eeq
Integrating both sides, we obtain
\begin{equation}
z^{b(2-a)}F(z) = -2\kappa r_c^a \rho_{ch} b \int z^{b(2-a)-1} (1-z)^{b(a-g)}dz 
\end{equation}
which is the same integral as in Eq.\eqref{beta_incompleta}, but $\alpha=b(2-a)$.
Then, we can written the solution $F(z)$ as
\begin{equation}
F(z) = \frac{2\kappa r^a_c \rho_{ch}}{(a-2)}{}~_2F_1\left(b(2-a), b(g-a); b(2-a)+1; z\right).
\end{equation}
Thus, returning to the Eq. \eqref{ansatzBTZ}, we obtain the metric for the $3D$ black hole sourced by dark matter
\beq\label{metricBTZ}
f(r)=2\left(\frac{r}{r_c}\right)^{-a}\frac{\kappa  \rho_{ch} r^2}{a-2} \, _2F_1\left(b(2-a), b(g-a); b(2-a)+1; -\left(\frac{r}{r_c}\right)^{1/b}\right)-M + \frac{r^2}{\ell^2},
\eeq

where we set $\Lambda=-1/\ell^2$.  

\begin{figure}[ht!]
    \centering
\includegraphics[width=0.6\textwidth]{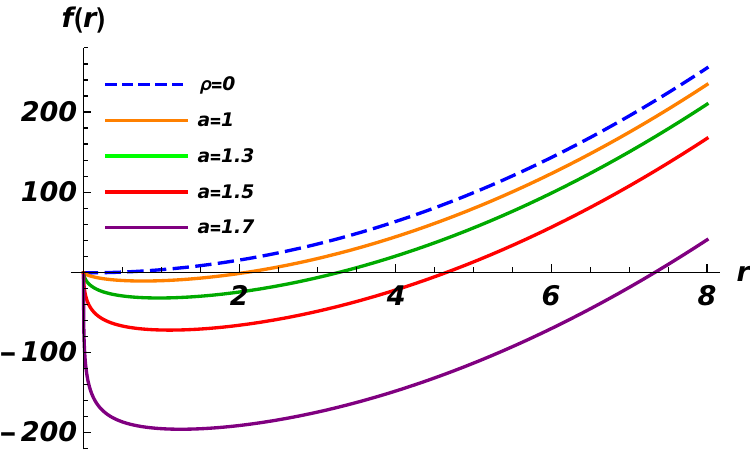}
        \caption{Metric function $f(r)$ for some values of $a$. We fixed $M=1$, $\ell=1/2$, $\rho_{ch}=6/5$, $g=7/2$, $b=2$, $r_c=3/2$ and $\kappa=8\pi$.}
\label{graficofBTZ}
\end{figure}  

The vacuum BTZ black hole is regular at the origin and has an event horizon at $r_h=\ell \sqrt{M}$. Figure \eqref{graficofBTZ} shows that regularity is preserved, and in this case, when generated by dark matter, the event horizon reaches a maximum as $a \rightarrow2$. As in the black string case, we analyze thermodynamic properties only in the interval $0<a<2$. Again, for $\rho_{ch}=0$, we recover the standard BTZ solution \cite{BTZ}.  

\subsection{Energy Conditions} 
With the energy-momentum tensor $T^{\mu}_{\ \nu} = (-\rho, -\rho, p_\theta)$, we can calculate the tangential  pressure, given by
\begin{equation}
p_\theta=\rho_{ch} \left(\frac{r}{r_c}\right)^{-a} \left[1+\left(\frac{r}{r_c}\right)^{1/b}\right]^{b(a-g)-1} \left[a-1+\left(g-1\right)\left(\frac{r}{r_c}\right)^{1/b} \right].\label{ptheta}
\end{equation}  

\begin{figure}[hb!]
%    \centering
    \includegraphics[width=0.60\linewidth]{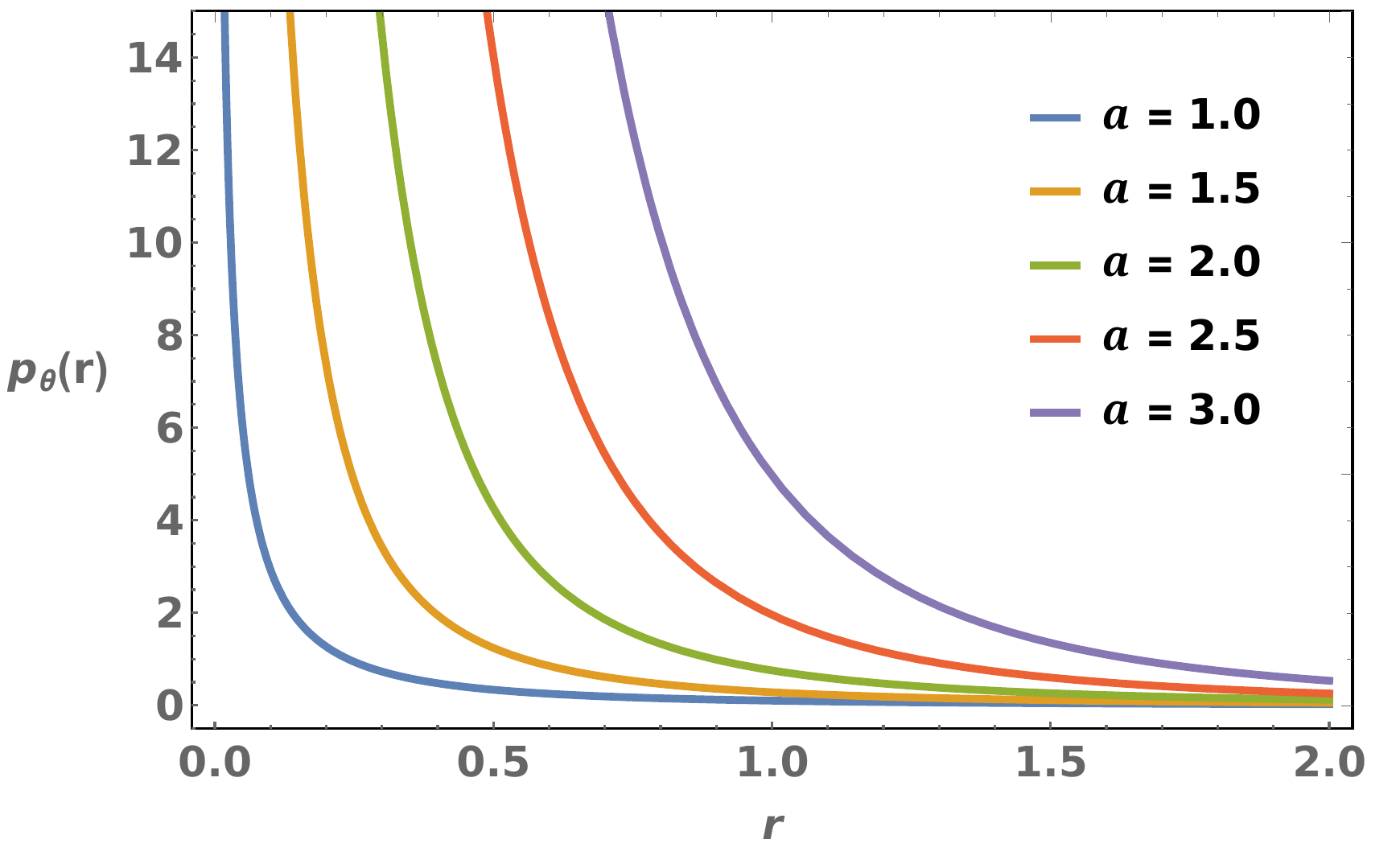}
    \caption{Tangential pressure $p_\theta$ for several values of $a$. We fixed $\mu=1$, $\ell=1/2$, $\rho_{ch}=6/5$, $r_c =3/2$, $g=7/2$, and $b=2$.}
    \label{figpL_BTZ}
\end{figure}

In this way, to evaluate the physical viability of the black hole sourced by the Dekel-Zhao profile, we examine the energy conditions derived from the effective energy-momentum tensor. Given that the field equations yield $p_r = -\rho$, the analysis primarily depends on the behavior of the tangential pressure $p_\theta$. Therefore, the Null Energy Condition (NEC) of the model defined by $\rho + p_i \ge 0$ is strictly satisfied throughout the spacetime since $p_r=-\rho$ and $p_\theta \ge 0$ [as shown in Eq. \eqref{ptheta} and Fig. \eqref{figpL_BTZ}], confirming that NEC is satisfied. 

The Weak Energy Condition (WEC) requires $\rho \ge 0$ and NEC. Both conditions are satisfied throughout the spacetime, confirming that local observers measure non-negative energy densities. This adherence to the WEC confirms that a physical observer following a timelike worldline would always measure a non-negative energy density, consistent with the dark matter distribution surrounding the $(2+1)$-dimensional gravity.

In $(2+1)$ dimensions, the Strong Energy Condition (SEC) requires $\rho + \sum p_i \geq 0$, which implies $\rho + p_r + p_\theta = \rho - \rho + p_\theta = p_\theta \ge 0$. As we can verify in Eq. \eqref{ptheta} and it is shown in Fig. \eqref{figpL_BTZ}, the tangential pressure $p_\theta$ remains non-negative for all physical values of the inner slope parameter $a$, for this choice of the parameters. Therefore, SEC is satisfied, indicating that the dark matter profile does not introduce exotic repulsive effects that violate classical gravitational behavior.

%Differently of the black string case, in the BTZ one the SEC is never violated because $p_L \ge 0$ for all values of the inner slope parameter $a$. Nevertheless, the dominant energy condition does not holds for all values of $a$, indicating that the dark matter cannot be interpreted as a causal fluid depending on the value of $a$. 

%Dominant Energy Condition (DEC) requires $\rho \geq |p_i|$. Since $|p_r| = \rho$ and the tangential pressure $p_\theta$ does not exceed the energy density in the regions of interest, the DEC is also preserved, ensuring the causality of the solution."

\begin{figure}[hb!]
%    \centering
    \includegraphics[width=0.55\linewidth]{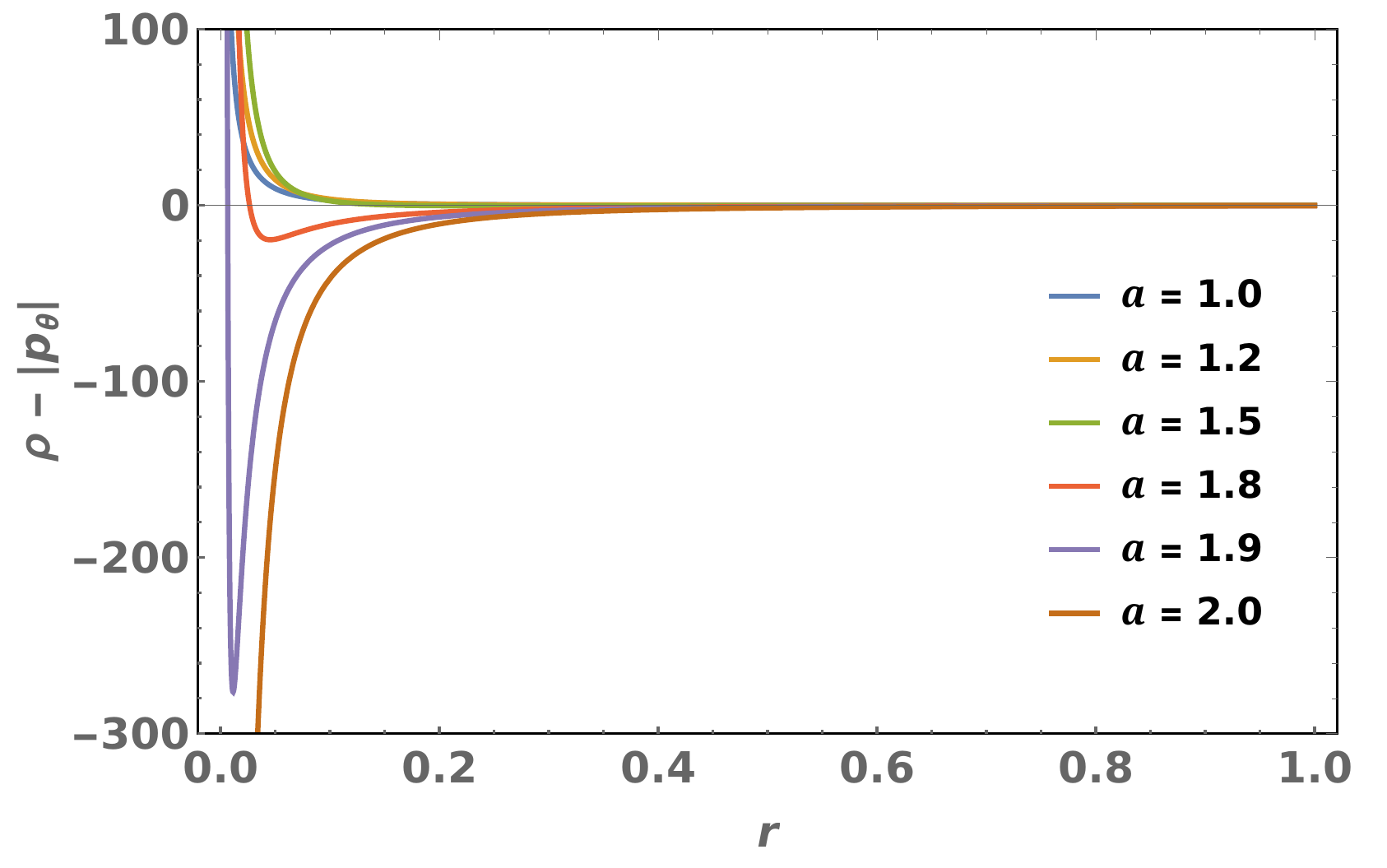}
    \includegraphics[width=0.40\linewidth]{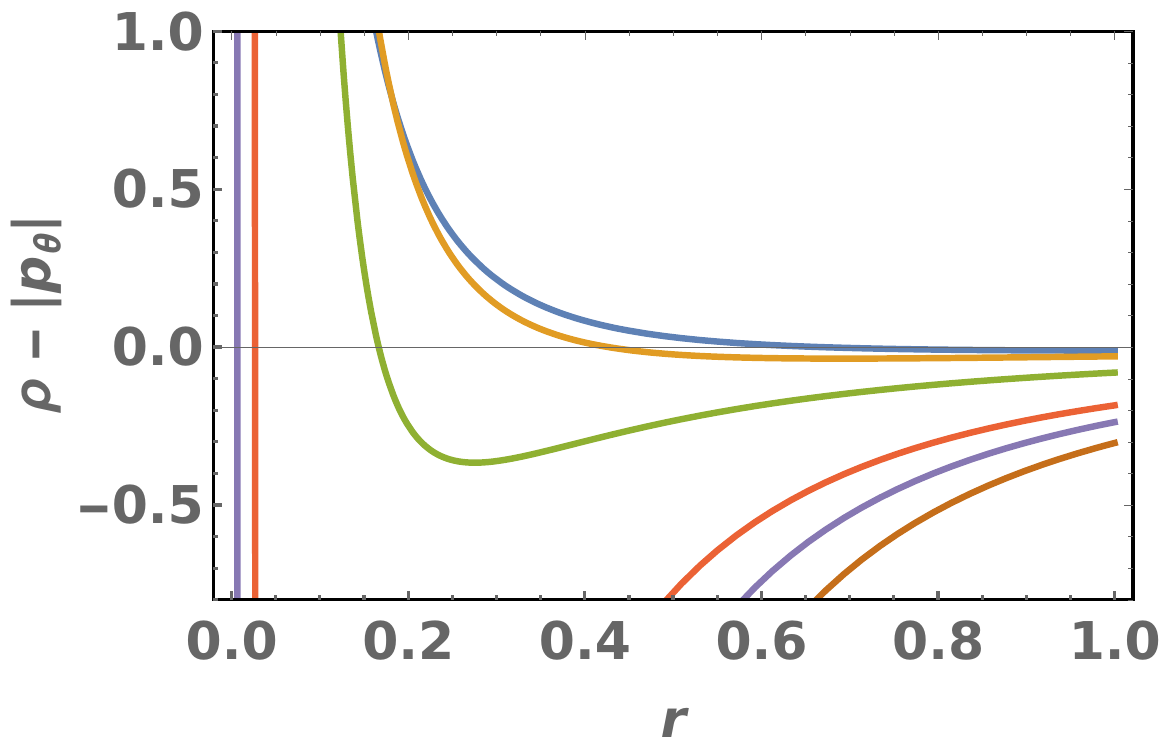}
    \caption{(Left) Dominant Energy Conditions (DEC) $\rho - |p_\theta| \geq 0$ as a function of radial coordinate $r$ for different values of the inner slope parameter $a$. (Right) Zoomed view showing the detailed behavior of DEC for the same values of $a$. We fixed $\mu=1$, $\ell=1/2$, $\rho_{ch}=6/5$, $g=7/2$, $b=2$, and $r_c =3/2$.}
    \label{DEC_BTZ}
\end{figure}

{The Dominant Energy Conditions (DEC) requires $\rho \geq |p_i|$. While this holds for the radial pressure, our numerical results show that $|p_\theta| > \rho$ within a specific range of the radial coordinate $r$ for all analyzed values of $a$ [Fig. \eqref{DEC_BTZ}]. Consequently, the DEC is violated. This suggests that the Dekel-Zhao profile imposes an extreme stiffness on the $(2+1)$-dimensional spacetime, where the transverse stresses necessary to support the configuration exceed the local energy density. 

As shown previously, the $3D$ black hole sourced by dark matter does not possess horizons for $a \geq 2$, and this behavior is directly associated with the violation of the dominant energy condition. We also observe that the region where the dominant energy condition is satisfied becomes progressively smaller as the parameter $a$ increases, as shown in  Fig.\eqref{DEC_BTZ}. This indicates that $a$ controls the transition between a physically acceptable black hole solution and an exotic configuration, where the dark matter source can no longer be interpreted as a causal fluid.
}
%the energy flux remains causal and that the dark matter source can be interpreted as a physically acceptable anisotropic fluid
%\begin{enumerate}
%\item \textbf{Null Energy Condition (NEC)}: $\rho +p_i\geq0$. \\
%
%\item \textbf{Weak Energy Condition (WEC)}: $\rho \geq0, \rho +p_i\geq0$. \\%
%
%\item \textbf{Strong Energy Condition (SEC)}: $\rho+p_L\geq0, p_L\geq0$. \\
%
%\item \textbf{Dominant Energy Condition (DEC)}: $\rho\geq |p_i|$.\\
%\end{enumerate}
%\begin{figure}[hb!]
%    \centering
  %  \includegraphics[width=0.60\linewidth]{pL_BTZ.pdf}
%     \includegraphics[width=0.35\linewidth]{grafico_pL_inset.pdf}
  %  \caption{Lateral pressure $p_L$ for several values of $a$. We fixed $\mu=1$, $\ell=1/2$, $\rho_{ch}=6/5$, $g=7/2$, $b=2$, and $\kappa=8\pi$.}
   % \label{figpL_BTZ}
%\end{figure}
\subsection{Curvature Invariants}  
\begin{figure}[ht!]
    \centering
\includegraphics[width=0.6\textwidth]{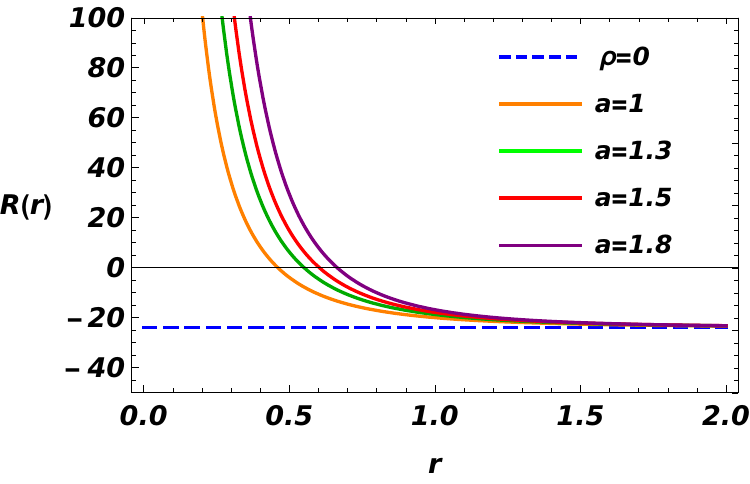}
\caption{Ricci scalar for some values of $a$. We fixed $M=1, \ell=1/2, \rho_{ch}=6/5, g=7/2, b=2,r_c=3/2 \ \text{and} \ \kappa=8\pi$.}
\label{ricciBTZ}
\end{figure}

Again, to obtain information about the singularities, we will study the curvature invariants. The Ricci scalar is  
\beq
R(r) = -\frac{6}{\ell^2} - 2 \kappa \rho_{ch} \left(\frac{r}{r_c}\right)^{-a} \Biggl\{ a + g \left(\frac{r}{r_c}\right)^{1/b} \!- 3 \left[ \left(\frac{r}{r_c}\right)^{1/b} \!+ 1 \right] \Biggr\} \left[ \left(\frac{r}{r_c}\right)^{1/b} \!+ 1 \right]^{b(a-g)-1}
\eeq  
For large $r$, it approaches the vacuum solution Fig.\eqref{ricciBTZ}. As expected, for $\rho_{ch}=0$, we recover the constant Ricci scalar \cite{BTZ}. Expanding for small $r$:  
\beq
R(r)\approx -\frac{6}{\ell ^2}-2 \kappa  \rho_{ch} \left(\frac{r}{r_c}\right)^{-a} \left[a+g \left(\frac{r}{r_c}\right)^{1/b}-3\right].
\eeq  
Thus, for $a>0$, the singularity is enhanced; for $a=0$, the spacetime becomes regular.  

The squared Ricci scalar is  
\bea
R^{\mu\nu}R_{\mu\nu}&=&\frac{12}{\ell ^4} +2 \kappa ^2 \rho_{ch}^2 \left(\frac{r}{r_c}\right)^{-2 a} \Bigg\lbrace a^2+2 a g \left(\frac{r}{r_c}\right)^{1/b}-4 a \left[\left(\frac{r}{r_c}\right)^{1/b}+1\right] +g^2 \left(\frac{r}{r_c}\right)^{2/b}\nonumber\\
&-&4 g \left[\left(\frac{r}{r_c}\right)^{1/b}+1\right] \left(\frac{r}{r_c}\right)^{1/b} +6 \left[\left(\frac{r}{r_c}\right)^{1/b}+1\right]^2\Bigg\rbrace\left[\left(\frac{r}{r_c}\right)^{1/b}+1\right]^{2 a b-2 b g-2} \nonumber\\
&+&\frac{8 \kappa \rho_{ch}}{\ell ^2} \left(\frac{r}{r_c}\right)^{\!-a}\left[\left(\frac{r}{r_c}\right)^{1/b}\!+1\right]^{ab-bg-1} \Bigg\lbrace a+g \left(\frac{r}{r_c}\right)^{1/b}\!\!-3 \left[\left(\frac{r}{r_c}\right)^{1/b}\!+1\right]\Bigg\rbrace,
\eea
that with $\rho_{ch}=0$, $R^{\mu\nu}R_{\mu\nu}=12/\ell^4$. For $r/r_c \ll 1$, we have
\beq
R^{\mu\nu}R_{\mu\nu} \approx 2 \kappa ^2 \rho_{ch}^2 \left(\frac{r}{r_c}\right)^{-2a}\left[a^2 +2g(a-2)\left(\frac{r}{r_c}\right)^{1/b}+g^2\left(\frac{r}{r_c}\right)^{2/b} -4a +6 \right]
\eeq
where for $a \leq 0$, $R^{\mu\nu}R_{\mu\nu}$ becomes regular, while for $a>0$, we have a singularity that becomes stronger as $a$ increases.

Finally, the Kretschmann scalar is  
\bea
K(r) &=& \frac{12}{\ell^4} 
+\frac{8 \kappa \rho_{ch}}{\ell^2} \left( \frac{r}{r_c} \right)^{-a}
\left[ 1 + \left( \frac{r}{r_c} \right)^{1/b} \right]^{-1 + ab - bg} 
\left[ -3 + a + (g-3) \left( \frac{r}{r_c} \right)^{1/b} \right] \nonumber\\
&+&4 \kappa^2 \rho_{ch}^2 \left[ 1 + \left( \frac{r}{r_c} \right)^{1/b} \right]^{-2 + 2ab - 2bg} 
\Bigg[ 3 - 2a + a^2 + 2 \left( 3 + a(g-1) - g \right) \left( \frac{r}{r_c} \right)^{1/b} \nonumber\\
&+& \left( 3 + g(g-2) \right) \left( \frac{r}{r_c} \right)^{2/b} \Bigg] 
\left( \frac{r}{r_c} \right)^{-2a}.
\eea

For $r/r_c \ll 1$: 
{\small
\bea
K(r) &\approx& 4 \kappa^2 \rho_{ch}^2 
\left( \frac{r}{r_c} \right)^{\!-2a} 
\Bigg\lbrace a^2 + 2 \left[ a(g-1) - g + 3 \right] \left( \frac{r}{r_c} \right)^{\tfrac{1}{b}}\! - 2a + \left[ (g-2)g + 3 \right] \left( \frac{r}{r_c} \right)^{\tfrac{2}{b}} \!+ 3 \Bigg\rbrace \nonumber\\
&+& \frac{8 \kappa \rho_{ch}}{\ell^2} 
\left( \frac{r}{r_c} \right)^{-a} 
\left[ a + g \left( \frac{r}{r_c} \right)^{1/b} - 3 \right] 
+ \frac{12}{\ell^4}.
\eea 
}

\begin{figure}[ht!]
    \centering
    \begin{subfigure}{0.48\textwidth}
        \centering
        \includegraphics[width=\linewidth]{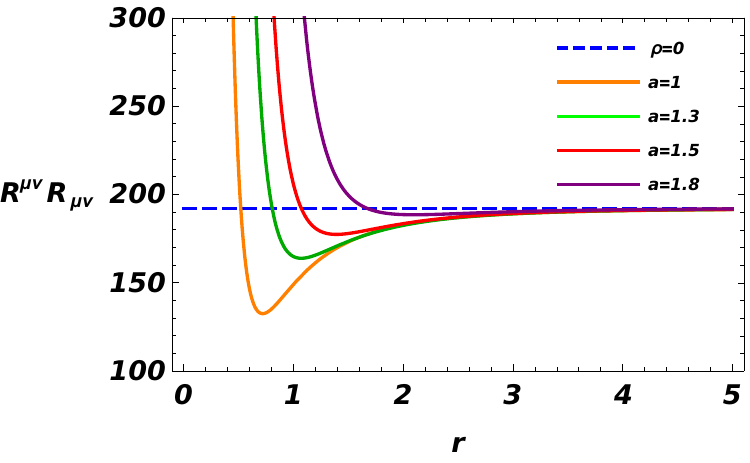}
        \label{ricci2BTZ}
    \end{subfigure}
   \hspace{0.02\textwidth}
    \begin{subfigure}{0.45\textwidth}
        \centering
        \includegraphics[width=\linewidth]{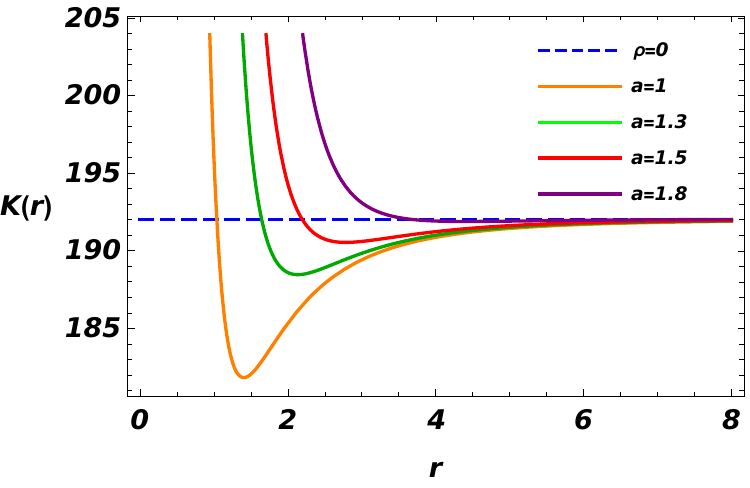}
        \label{kretBTZ}
    \end{subfigure}
    \caption{$R^{\mu\nu}R_{\mu\nu}$ and $K(r)$, respectively, for some values of $a$. We fixed $M=1$, $\ell=1/2$, $\rho_{ch}=6/5$, $g=7/2$, $b=2$, $r_c=3/2$ and $\kappa=8\pi$.}
    \label{KRIcci2}
\end{figure}

Due to the dark matter profile, the Kretschmann scalar becomes singular near the origin, of order $r^{-2a}$, which increases with $a$. For $a\le 0$, it becomes regular. In the black string case, the Kretschmann scalar in vacuum is already singular, and dark matter strengthens this singularity. In the $3$D black hole, which is normally of constant curvature (BTZ), dark matter induces a singular Kretschmann scalar. The plots of $R^{\mu\nu}R_{\mu\nu}$ and $K(r)$ are shown in Fig. \eqref{KRIcci2}.

\subsection{Thermodynamics}  
In this section, we study the thermodynamics of the $3$D black hole surrounded by dark matter. First, we analyze the Hawking temperature. 

\begin{figure}[ht!]
    \centering
\includegraphics[width=0.6\textwidth]{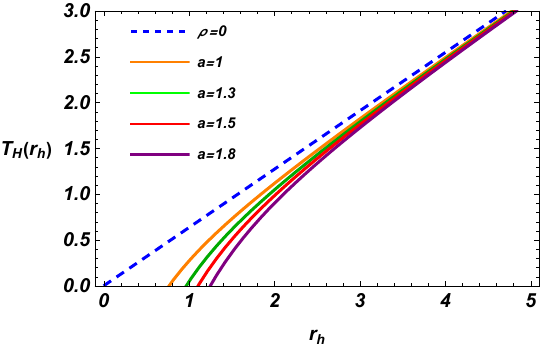}
        \caption{Hawking temperature for BTZ black hole for some values of $a$. We fixed $ \ell=1/2, \rho_{ch}=6/5, g=7/2, r_c=3/2  \ \text{and} \ b=2$.}
\label{ThawkingBTZ}
\end{figure}

Using \eqref{metricBTZ} in \eqref{thawking}, we obtain the Hawking temperature as a function of the event horizon:
\beq
T_H=\frac{r_h}{4\pi}\Bigg\lbrace \frac{2}{\ell^{2}} - 2\kappa \rho_{ch}\left(\frac{r_h}{r_c}\right)^{-a}\left[\left(\frac{r_h}{r_c}\right)^{\tfrac{1}{b}} + 1\right]^{\,b(a-g)}\Bigg\rbrace,
\eeq
where for $\rho_{ch}=0$, we recover the usual solution $T_H=r_h /2\pi \ell^2$ \cite{BTZ}. In vacuum, upon reaching the minimum temperature, black hole evaporates completely $(r_h^{min}=0)$. However, in the presence of dark matter there is a remnant mass density. It is not possible to obtain an analytical expression for $r_h^{min}$. Since negative Hawking temperature is not physically meaningful, the existence of black holes is restricted to the region $r_h \ge r_h^{min}$. The plot of the Hawking temperature as a function of the horizon radius is shown in Fig.\eqref{ThawkingBTZ}.

Calculating the entropy $dM/T_H$, we obtain $S=4\pi r_h$, which is the usual result for the entropy of the BTZ black hole \cite{BTZ}, being proportional to the horizon length, since we are in $(2+1)$ dimensions.

\begin{figure}[ht!]
    \centering
   % \begin{subfigure}{0.44\textwidth}
        \centering
        \includegraphics[width=0.5 \textwidth]{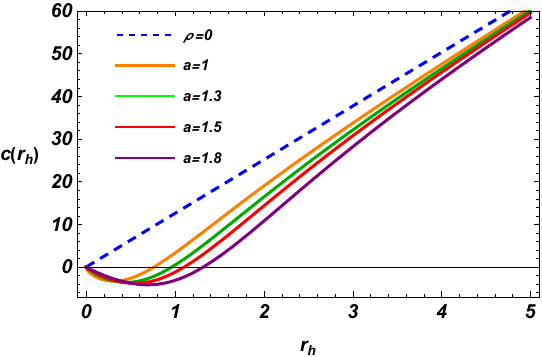}
 %       \label{heatcBTZ}
    %\end{subfigure}
  %  \hspace{0.05\textwidth}
    %\begin{subfigure}{0.44\textwidth}
       %\centering
       % \includegraphics[width=\linewidth]{heatc0BTZ.pdf}
     %   \label{heatc0BTZ}
  %  \end{subfigure}
    \caption{Heat capacity for some values of $a$. We fixed  $\ell=1/2$, $\rho_{ch}=6/5$, $g=7/2$, $r_c=3/2$ and $b=2$.}
    \label{heatcBTZ}
\end{figure}

Now, analyzing the heat capacity, $c=dM/dT_H$, we can write the mass as a function of the horizon radius,
\beq
M(r_h) = \frac{r^2_h}{\ell^2}+\left(\frac{r_h}{r_c}\right)^{-a}\frac{\kappa  \rho_{ch} r^2_h}{a-2} \, _2F_1\left((2-a) b,b (g-a);(2-a) b+1;-\left(\frac{r_h}{r_c}\right)^{1/b}\right).
\eeq
Thus, we can write $c=\frac{dM}{d r_h} \frac{d r_h}{dT_H}$. In this way, we obtain the expression for the heat capacity
\beq
c= \frac{4 \pi  r_h \left[\left(\frac{r}{r_c}\right)^{1/b}+1\right] \Bigg\lbrace\left(\frac{r_h}{r_c}\right)^a-\kappa \rho_{ch} \ell ^2 \left[\left(\frac{r_h}{r_c}\right)^{1/b}+1\right]^{b (a-g)}\Bigg\rbrace}{\kappa \rho_{ch} \ell ^2 \left[\left(\frac{r_h}{r_c}\right)^{1/b}+1\right]^{b (a-g)} \left[a-1+(g-1)\left(\frac{r_h}{r_c}\right)^{1/b}\right]+\left(\frac{r_h}{r_c}\right)^a \left[\left(\frac{r_h}{r_c}\right)^{1/b}+1\right]}.
\eeq
Taking $\rho_{ch}=0$, we get $c=4\pi r_h$, that is, the heat capacity in vacuum equals the entropy, a result that occurs only for the BTZ black hole. In Fig.\eqref{heatcBTZ} we see the behavior of the heat capacity. The regions with negative heat capacity are not physically acceptable since they occur in regions with negative temperatures. %We can observe that the BTZ black hole generated by dark matter, depending on the choice of $a$, is locally stable and does not exhibit a phase transition. 

%For $a=0$, We observe in Fig.\eqref{heatcBTZ} two phase transitions: a second-order transition at the critical point $r_h \approx 0.0533$, and a first-order transition at $r_h \approx 0.167$.

\begin{figure}[ht!]
    \centering
   % \begin{subfigure}{0.44\textwidth}
        \centering
        \includegraphics[width=0.5\linewidth]{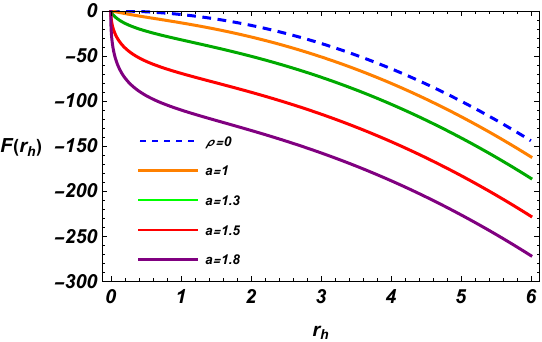}
   % \end{subfigure}
   % \hspace{0.05\textwidth}
   % \begin{subfigure}{0.44\textwidth}
        %\centering
      %  \includegraphics[width=\linewidth]{freeenergy0BTZ.pdf}
     %   \label{freeenergy0BTZ}
  %  \end{subfigure}
    \caption{Free energy for some values of $a$. We fixed  $\ell=1/2$, $\rho_{ch}=6/5$, $g=7/2$, $r_c=3/2$ and  $b=2$.}
    \label{freeenergyBTZ}
\end{figure}

Finally, analyzing the Helmholtz free energy, $F = M - T_H S$, we obtain
\bea
F&=&-\frac{r^2}{(a-2)\ell^2}\left(\frac{r}{r_c} \right)^{-a}\Bigg\lbrace \left[\left(\frac{r_h}{r_c}\right)^{a} - \left(1+\left(\frac{r_h}{r_c}\right)^{1/b}\right)^{b(a-g)}2 \ell^2 \kappa \rho_{ch}\right] \nonumber\\
&-&2\ell^2 \kappa \rho_{ch}\, _2F_1\left((2-a) b,b (g-a);(2-a) b+1;-\left(\frac{r_h}{r_c}\right)^{1/b}\right) \Bigg\rbrace,
\eea
Thus, we can see that the $3$D black hole remains thermodynamically stable in the presence of dark matter, since the free energy $F$ is negative throughout the physical parameter range Fig.\eqref{freeenergyBTZ}.

\subsection{Geodesics}
Now, we will analyze the equations of motion for particles near a $3$D black hole sourced by dark matter. The geodesic equations are the same as those for the black string \eqref{geodesics}, but without the $z$ coordinate. With the conserved quantities $\dot\phi=L/r^2$ and $\dot t=E/f(r)$, we have the following effective potential
\beq \label{veff}
V_{eff}(r)=f(r)\left(\frac{L^2}{r^2}-\epsilon \right).
\eeq
It is known that circular photon orbits in the BTZ black hole are constrained by the event horizon $r_0 \le r_h$, which is a non-physical result, since all particles crossing the horizon are inevitably drawn to $r=0$. Now, when sourced by dark matter, is it possible to have circular photon orbits?

\begin{figure}[ht!]
    \centering
\includegraphics[width=0.6\textwidth]{ 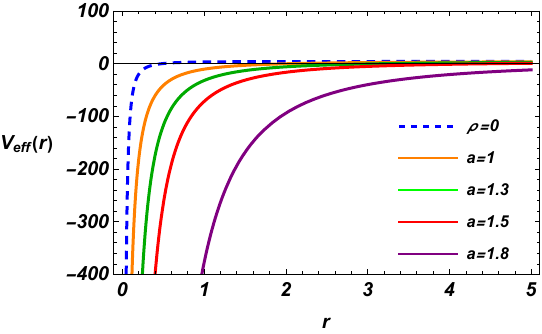}
        \caption{Effective potential for photons for 3D black hole for some values of $a$. We fixed $M=1, L=1, \ell=1/2, \rho_{ch}=6/5, g=7/2 \ \text{and} \ b=2.$}
\label{GeodesicaBTZ}
\end{figure}

Again, for circular photon orbits, we must impose the condition $V'_{eff}(r_0)=0$, with $r_0$ being the orbit radius. Substituting \eqref{metricBTZ} into \eqref{veff}, we have
\bea
V_{eff}(r)&=&2L^2\left(\frac{r}{r_c}\right)^{-a}\frac{\kappa  \rho_{ch} }{a-2} \, _2F_1\left((2-a) b,b (g-a);(2-a) b+1;-\left(\frac{r}{r_c}\right)^{1/b}\right)\nonumber\\
&-&\frac{ML^2}{r^2}+\frac{L^2}{\ell^2}.
\eea
whose derivative is
\bea
V'_{eff}(r)&=&\left(\frac{r}{r_c}\right)^{-a}\frac{4L^2\kappa \rho_{ch}}{(2-a)r} \, _2F_1\left((2-a) b,b (g-a);(2-a) b+1;-\left(\frac{r}{r_c}\right)^{1/b}\right)\nonumber\\
&+&\frac{2ML^2}{r^3} - \frac{2L^2\kappa\rho_{ch}}{r}\left[1+\left(\frac{r}{r_c}\right)^{1/b}\right]^{b(a-g)}
\eea
Again, we can see that there is no $r_0$ that satisfies $V_{eff}(r_0)=0$, and therefore, there are no circular photon orbits in the $3$D black hole sourced by dark matter, similarly to the BTZ solution \cite{BTZ}. The plot of the effective potential is shown in Fig.\eqref{GeodesicaBTZ}, where we considered some values of the parameter $a$.

\section{Conclusions}
In this study, we investigated the gravitational and thermodynamic implications of immersing black strings and lower-dimensional black holes into a Dekel-Zhao dark matter profile. A primary conclusion is that the DM distribution significantly reshapes the horizon structure. In both geometries, the event horizon expands with the inner slope parameter $a$ until a limit is reached where the horizon vanishes, resulting in a naked singularity.

The evaluation of the energy-momentum tensor provided substantial insights into the nature of these solutions. We demonstrated that the Strong Energy Condition (SEC) is satisfied, ensuring that the dark matter environment preserves the standard geodesic focusing characteristic of classical gravity. However, the noteworthy violation of the Dominant Energy Condition (DEC) in the $3$D case -- where the tangential pressure $p_\theta$ exceeds the energy density $\rho$ -- highlights the extreme relativistic stiffness required to sustain the DM halo in $(2+1)$ dimensions.

Furthermore, while the vacuum $(2+1)$ spacetime is characterized by constant curvature (BTZ), the introduction of the Dekel-Zhao profile induces a true curvature singularity at the origin. The interplay between this singularity and the vanishing of the horizon suggests that a sufficiently cuspy dark matter profile prevents the formation of a black hole, exposing the singularity to distant observers. These findings contribute to a deeper understanding of the cosmic censorship hypothesis in dark-matter-driven geometries and provide theoretical distinctions that could be relevant for future observational studies of non-standard black hole solutions.
\acknowledgments{The authors G.A. and M.S.C. would like to thank Conselho Nacional de Desenvolvimento Cient\'{i}fico e Tecnol\'{o}gico (CNPq) and Fundação Cearense de Apoio ao Desenvolvimento Científico e Tecnológico (FUNCAP) through PRONEM PNE0112- 00085.01.00/16, for the partial financial support. V.H.U.B. is supported by Coordena\c c\~{a}o de Aperfei\c coamento de Pessoal de N\'{i}vel Superior - Brasil (CAPES) Finance Code 001. R.R.L. is supported in part by Conselho Nacional de Desenvolvimento Cient\'{i}fico e Tecnol\'{o}gico (CNPq).
}
\bibliographystyle{apsrev4-1}
\bibliography{referencias}
\end{document}